\documentclass{IEEEtran}
\usepackage{amsfonts}
\usepackage{mathrsfs}
\usepackage{mathtools}
\usepackage{amssymb,amsmath}

\usepackage{algorithm}
\usepackage{algorithmic}
\usepackage{amsmath,amssymb,epsfig,graphics,subfigure}
\usepackage{theorem}
\usepackage{array,color}
\usepackage[compress]{cite}
\usepackage{tabularx}
\usepackage{multicol}
\usepackage{bm}
\usepackage{cite}
\usepackage{stfloats}
\usepackage{balance}

\newtheorem{remark}{Remark}
\newtheorem{lemma}{Lemma}

\newtheorem{definition}{Definition}

\allowdisplaybreaks

\begin{document}

\title{Matrix-Monotonic Optimization $-$ Part II: Multi-Variable Optimization}

\author{Chengwen Xing, \textsl{Member}, \textsl{IEEE}, Shuai Wang, \textsl{Member}, \textsl{IEEE},
 Sheng Chen, \textsl{Fellow}, \textsl{IEEE}, Shaodan Ma, \textsl{Member}, \textsl{IEEE}, H. Vincent Poor, \textsl{Fellow}, \textsl{IEEE}, and  Lajos Hanzo, \textsl{Fellow}, \textsl{IEEE}
 \thanks{This work was supported in part by the U.S. National Science Foundation under Grants CCF-0939370 and CCF-1513915.}
 \thanks{L. Hanzo would like to thank the ERC for his Advanced Fellow Award and the EPSRC for their financial support.}
\thanks{C.~Xing and S.~Wang are with School of Information and Electronics, Beijing
 Institute of Technology, Beijing 100081, China (E-mails: xingchengwen@gmail.com
 swang@bit.edu.cn)} %
\thanks{S.~Chen and L.~Hanzo are with School of Electronics and Computer Science,
 University of Southampton, U.K. (E-mails: sqc@ecs.soton.ac.uk, lh@ecs.soton.ac.uk).
 S. Chen is also with King Abdulaziz University, Jeddah, Saudi Arabia.} %
\thanks{S.~Ma is with the State Key Laboratory of Internet of Things for Smart City and Department of Electrical and Computer Engineering, University
 of Macau, Macao (E-mail: shaodanma@um.edu.mo).} %
\thanks{H.~V.~Poor is with Department of Electrical Engineering, Princeton University,
 Princeton, NJ 08544 USA (E-mail: poor@princeton.edu).} %
\vspace*{-5mm}
}

\maketitle

\begin{abstract}In contrast to Part I of this treatise~\cite{Matrixmonotonic} that focuses on the optimization problems associated with single matrix variables, in this paper, we investigate the application of the matrix-monotonic optimization framework in the optimization problems associated with multiple matrix variables. It is revealed that matrix-monotonic optimization still works even for multiple matrix-variate based optimization problems, provided that certain conditions are satisfied. Using this framework, the optimal structures of the matrix variables can be derived and the associated multiple matrix-variate optimization problems can be substantially simplified. In this paper several specific examples are given, which are essentially open problems. Firstly, we investigate multi-user multiple-input multiple-output (MU-MIMO) uplink communications under various power constraints. Using the proposed framework, the optimal structures of the precoding matrices at each user under various power constraints can be derived. Secondly, we considered the optimization of the signal compression matrices at each sensor under various power constraints in distributed sensor networks. Finally, we investigate the transceiver optimization for multi-hop amplify-and-forward (AF) MIMO relaying networks with imperfect channel state information (CSI) under various power constraints. At the end of this paper, several simulation results are given to demonstrate the accuracy of the proposed theoretical results.
\end{abstract}


\section{Motivations}\label{sect:intro}

The deployment of multi-antenna arrays opened a door to effectively exploit spatial resources to improve energy efficiency and spectrum efficiency \cite{Matrixmonotonic,Sugiura201201,Kadir2015,Sugiura201202,Shaoshi2015}. Meanwhile, the involved design variables are usually matrices instead of simple scalars \cite{JYang1994,Sampth02,Feiten2007}.
In order to solve the matrix-variate optimization problems for MIMO communications efficiently, the most widely used logic is first to derive the optimal structures of the matrix variables. Then based on the optimal structures, the considered optimization problems can be greatly simplified \cite{Yadav2014,Yao_etal2010,Jafar2004,Ding09,Pastore2012,ShiqiTSP2017,Zhang2008,Jafar2005}.

Matrix-monotonic optimization is an interesting framework that takes
advantage of monotonic property in positive semidefinite matrix set to
derive the optimal structures of optimization variables
\cite{Matrixmonotonic,XingTSP201901,XingTSP201501,XingJSAC202001,NewTSP}. In
  Part I \cite{Matrixmonotonic}, we focus our attention on
single-variable optimization problems. However, for many practical
optimization problems there are multiple matrix variates to
optimize. For example, in multi-user multiple-input multiple-output
(MU-MIMO) communication systems, the transceiver optimization
processes of the downlink and uplink involve multiple matrix
variables, namely the equalizer matrices and precoder matrices
\cite{Shi2011,Serbetli2004,Goldsmith2003b,WYu2007}. For multi-carrier MIMO
systems, in each subcarrier there is a precoder matrix and an
equalizer matrix \cite{XingTSP201501}. Moreover, in multi-hop
communications the forwarding matrix of each relay has to be
optimized~\cite{XingTSP2013,XingJSAC2012}.

This fact inspires us to take a further step and to investigate the
optimization problems hinging on multiple matrix-variables.  Generally
speaking, solving an optimization problem having multiple
matrix-variables is more challenging than its single matrix-variable counterpart. How to solve this
kind of optimization problems has attracted substantial attention both
across the wireless communication and signal processing research
communities~\cite{WYu2007,Goldsmith2003b,Serbetli2004}. In contrast to
single matrix-variable optimizations, for multiple matrix-variable
optimization in most cases it is impossible to derive the optimal
solutions in closed-form.  Iterative optimization algorithms or
alternating optimization algorithms have neeb widely used to solve
this kind of optimization
problems~\cite{Serbetli2004,Shi2011,JFang2013,XingTSP2013,XingJSAC2012}.
Unfortunately, there is no general-purpose mathematical tool or
framework that can cover all the kinds of optimization problems. In
some cases, similar to the single-variable case, for multiple
matrix-variable optimization first the optimal structures of the matrix
variables have to be derived, based on which the optimization can
be significantly simplified and the corresponding convergence rate can
be substantially improved.

In this paper, we investigate in detail, how to exploit the hidden
monotonicity in positive semidefinite matrix fields to derive the
optimal structures of the multiple matrix variables. Based on the
optimal structures, the optimizations of multiple matrix variables can
be significantly simplified. In our work, it is revealed that for many
optimization problems associated with multiple matrix variables, the
matrix-monotonic optimization framework still works.  We also would
like to point out that the authors of~\cite{XingTSP201501} also
investigate how to apply matrix-monotonic optimization to optimization
problems associated with multiple matrix variables. However, it is
worth highlighting that the previous contribution~\cite{XingTSP201501}
only considers a simple sum power constraint. By contrast, our work in
this paper is significantly different from that in
\cite{XingTSP201501}, since here diverse power constraints are taken
into account, such as the multiple weighted power constraints
of~\cite{NewTSP}, the shaping constraint
of~\cite{Dai2012,XingTSP201502} and so on. Additionally, more
scenarios are also taken into account. Furthermore, in addition to the
multi-hop systems investigated in~\cite{XingTSP201501}, in this paper,
the MU-MIMO uplink and distributed sensor networks are also
considered.


The main contributions of this paper are enumerated in the following. These contributions distinguish our work from the existing related works.

\begin{itemize}
\item Firstly, we investigate precoder optimization in the uplink
  of MU-MIMO communications under three different power constraints,
  namely the shaping constraint, joint power constraint and multiple
  weighted power constraints. Based on the matrix-monotonic
  optimization framework, the optimal structures of the matrix
  variables can be derived. Then the optimization can be substantially
  simplified and can be efficiently solved by an iterative
  algorithm. In each iteration based on the optimal structure, the
  optimal solutions of the remaining variables are standard
  water-filling solutions. We cover the precoder optimization under
  per-antenna power constraint as its special cases.

\item Secondly, we investigate the signal compression matrix
  optimization problem in a distributed sensor network under the above
  three power constraints. For this data fusion optimization, there
  exist correlations between the signals transmitted from different
  sensors. This makes the corresponding optimization problem
  significantly different from that in the MU-MIMO uplink. Moreover,
  in contrast to~\cite{JFang2013}, where at each sensor only the sum
  power constraint is considered, in our work more general power
  constraints are taken into account, namely the shaping constraint,
  joint power constraint and multiple weighted power constraints. This
  is our main contribution.  Based on the matrix-monotonic
  optimization framework, the optimal structures of the compression
  matrices can be derived and the optimal solutions of the remaining
  vectors are found to correspond to  water-filling solutions.

\item Thirdly, we investigate the robust transceiver optimization
  problem of multi-hop amplify-and-forward (AF) cooperative MIMO
  networks, including both linear and nonlinear transceivers. For the
  linear transceivers, various kinds of performance metrics are taken
  into account, namely the additively Schur-convex and Schur-concave
  scenarios~\cite{XingTSP2013,XingJSAC2012,Jorswieck07}. On the other
  hand, for nonlinear transceivers, both decision feedback equalizers
  (DFE) and Tomlinson-Harashima precoding (THP) are investigated.  In
  contrast to~\cite{XingTSP201502,XingTSP201601}, various power
  constraints are taken into account in the robust transceiver
  optimization instead of the simple sum power constraint. Based on
  the proposed framework, the optimal structures of the robust
  transceiver design can be derived, based on which the robust
  transceiver optimization can be efficiently solved. Hence our
  contribution fills a void in the robust transceiver design
  literature of multi-hop AF MIMO systems under multiple weighted
  power constraints.
\end{itemize}

 The remainder of this paper is organized as follows. In Section II, the basic properties of the framework on matrix-monotonic optimizations are given first.
 Following that, the MU-MIMO uplink is investigated in
 Section~\ref{S6}. Compression matrix optimization for distributed sensor networks is
 discussed based on matrix-monotonic optimization in Section~\ref{S7}. In Section~\ref{S8},
 robust transceiver optimization is proposed for multi-hop AF MIMO relaying networks separately under shaping
 constraints, joint power constraints and multiple weighted power constraints. Several numerical results are given in Section~\ref{S9},
 Finally, the conclusions are drawn in Section~\ref{S10}.

\noindent \textbf{Notation:} To be consistent with our Part I work \cite{Matrixmonotonic}, the following notations and definitions are used throughout this paper. The symbols $\bm{Z}^{\rm H}$, $\bm{Z}^{\rm{T}}$, ${\rm Tr}(\bm{Z})$ and $|\bm{Z}|$ denote
 the Hermitian transpose, transpose, trace and determinant of matrix $\bm{Z}$, respectively. The matrix
 $\bm{Z}^{\frac{1}{2}}$ is the Hermitian square root of a positive semidefinite matrix $\bm{Z}$, which is also a positive
 semidefinite matrix. For an $N\times N$ matrix ${\bm{Z}}$, the vector  ${\boldsymbol \lambda}({\bm{Z}})$ is defined as ${\boldsymbol \lambda}({\bm{Z}})=[\lambda_1({\bm{Z}}),\cdots,\lambda_N({\bm{Z}})]^{\rm{T}}$  where $\lambda_i(\bm{Z})$ denotes the $i$th largest eigenvalue of $\bm{Z}$.
 The symbol $[\bm{Z}]_{i,j}$ denotes the $i$th-row and $j$th-column element.
 On the other hand, the symbol $\bm{d}(\bm{Z})$ denotes the vector consisting of the diagonal elements of $\bm{Z}$. The
 identity matrix is denoted by $\bm{I}$. In this paper, $\bm{\Lambda}$ always represents a diagonal matrix, and $\bm{\Lambda}\searrow$ and $\bm{\Lambda}\nearrow $ represent a rectangular or
 square diagonal matrix with the diagonal elements in descending order and ascending order,
 respectively.

\section{Fundamentals of Matrix-Monotonic Optimization}\label{S2}

In this paper, we investigate a real valued optimization problem with multiple complex matrix variables $\{\bm{X}_k\}_{k=1}^K$ which is generally
 formulated as
\begin{align}\label{general_optimization} 
 \textbf{Opt.\,1.1:}  & \min\limits_{\{\bm{X}_k\}_{k=1}^K}  f_0(\{\bm{X}_k\}_{k=1}^K) , \nonumber \\
  & \ \ \ {\rm{s.t.}} \ \ \ \  \psi_{k,i}(\bm{X}_k)\le 0, \nonumber \\
  & \ \ \ \ \ \ \ \ \ \ \ 1\le k\le K, 1\le i\le I_k,
\end{align}
where $\psi_{k,i}(\cdot )$, $1\le k \le K$,
 $1\le i\le I_k$, denotes the constraint functions. Similar to the single-variate matrix-monotonic optimization investigated in Part I \cite{Matrixmonotonic}, all constraints considered in this paper are right unitarily invariant, i.e., for arbitrary unitary matrices ${\bm{Q}}_{\bm{X}_k}$'s,
 \begin{align}
 \psi_{k,i}\left(\bm{X}_k{\bm{Q}}_{\bm{X}_k}\right)=\psi_{k,i}\left(\bm{X}_k\right).
 \end{align}In the following, several specific power constraints are given. The general power constraint model is one of the main contributions of this work.

\subsection{The Constraints on Multiple Matrix Variables}

 The simplest constraint is sum power constraint, i.e., the sum power across all transmit antennas is smaller than a predefined threshold. For example, in MU-MIMO uplink communications, each mobile terminal has a sum power constraint such as
\begin{align}\label{eq2}
{\rm Tr}\big(\bm{X}_k\bm{X}_k^{\rm H}\big)\le P_k.
\end{align} It is obvious that the sum power constraint is right unitarily invariant. Moreover, in order to constrain the fluctuation of the eigenvalues of $\bm{X}_k\bm{X}_k^{\rm H}$,
the following joint power constraint will be used \cite{XingTSP201502,Dai2012}
 \begin{align}\label{eq9}
 &  {\rm Tr}\big(\bm{X}_k\bm{X}_k^{\rm H}\big) \le P_k, \ \
  \bm{X}_k\bm{X}_k^{\rm H} \preceq \tau_k \bm{I}.
\end{align} The difference between the sum power constraint and the joint power constraint is that there is an additional maximum eigenvalue constraint. It is obvious that the joint power constraint is right unitarily invariant.

From the circuit viewpoint, each amplifier is connected to one distinct antenna.
It is not reasonable to use the sum power as a constraint as the powers cannot be shared between different antennas.
In other words, the individual power constraint or per-antenna power constraint is more practical, which is formulated as \cite{XingTSP201601,Mai2011,WYu2007}
\begin{align}\label{eq3}
 \big[\bm{X}_k\bm{X}_k^{\rm H}\big]_{n,n}\le P_{k,n}, \ \ n=1,\cdots ,N.
\end{align} The per-antenna power constraint is also right unitarily invariant. It is worth highlighting that the per-antenna power constraint cannot include the sum power constraint as its special case.

In order to build a more general constraint model including more specific power constraints as its special cases, multiple weighted power constraints are given in the following \cite{Matrixmonotonic,NewTSP}
\begin{align}\label{eq4}
 {\rm Tr}\big(\bm{\Omega}_{k,i}\bm{X}_k\bm{X}_k^{\rm H}\big) \le P_{k,i},
  \ \ i=1,\cdots ,I_k ,
\end{align}
 where $I_k$ is the number of weighted power constraints for the $k$th variable ${\bm{X}}_k$.
 The positive semidefinite matrices $\bm{\Omega}_{k,i}$'s are weighting matrices. The multiple weighted power constraints are right unitarily invariant as well.

Finally, in order to constrain the transmit signals to be in a desired region, shaping constraint can be used. Shaping constraint is a constraint on the covariance matrix of transmitted signals. Specifically, the shaping constraint on a matrix variable $ \bm{X}_k$ is defined as  \cite{XingTSP201502,Palomar2004}
\begin{align}\label{eq5}
 \bm{X}_k\bm{X}_k^{\rm H} \preceq \bm{R}_{{\rm s}_k},
\end{align} where $\bm{R}_{{\rm s}_k}$ is the desired signal shaping matrix \cite{XingTSP201502,Palomar2004}.
The shaping constraint (\ref{eq5}) is right unitarily invariant as well. Under these power constraints, in the following we give the properties which are the basis of application of the framework of matrix-monotonic optimization.

From a mathematical perspective, the more complicated power
constraints will significantly change the feasible set compared to
that of the sum power constraint. This is because the sum power
constraint is both right unitarily invariant and left unitarily
invariant, however the general power constraints are only right
unitarily invariant. In other words, the symmetry of sum power
constraint does not exist for the general power constraints such as
multiple weighted power constraints.  It is clear that under multiple
weighted power constraints the extreme values and the optimal
solutions are significantly different from that under the sum power
constraint.  The multiple weighted power constraints model also
includes the sum power constraint model as its special case. Note that
the sum power constraint model is not a special case of the
per-antenna power constraint model. One model can include two
different constraint models as its special cases. This is also an
advantage of the multiple weighted power constraints model.

\subsection{Matrix-Monotonic Properties}

The framework of matrix-monotonic optimization aims at exploiting the monotonicity in positive semidefinite field to derive the optimal structures of the matrix variates. As the constraints in \textbf{Opt.\,1.1} are right unitarily invariant,
defining $\bm{X}_k=\bm{F}_k{\bm{Q}}_{\bm{X}_k}$  \textbf{Opt.\,1.1} is equivalent to  the following optimization problem
\begin{align}
  \textbf{Opt.\,1.2:} \ \ \min\limits_{\{\bm{F}_k,{\bm{Q}}_{\bm{X}_k}\}_{k=1}^K}  \ \ &f_0(\{\bm{F}_k{\bm{Q}}_{\bm{X}_k}\}_{k=1}^K) , \nonumber \\
 {\rm{s.t.}} \ \ \ \  &\psi_{k,i}(\bm{F}_k)\le 0, \nonumber \\
 &1\le k\le K, 1\le i\le I_k
\end{align} In our work,  \textbf{Opt.\,1.2:} satisfies the following properties.

For the $k$th optimal unitary matrix
  ${\bm{Q}}_{\bm{X}_k}$, the objective function in \textbf{Opt.\,1.2}
can be transferred into a function of
${\boldsymbol\lambda}(\bm{F}_k^{\rm H}\bm{\Pi}_k\bm{F}_k)$ i.e.,
\begin{align}
f_0(\{\bm{F}_k{\bm{Q}}_{\bm{X}_k}\}_{k=1}^K)=g_{0,k}({\boldsymbol\lambda}(\bm{F}_k^{\rm H}\bm{\Pi}_k\bm{F}_k)), \text{for} \ k=1,\cdots,K
\end{align} with $g_{0,k}({\boldsymbol\lambda}(\bm{F}_k^{\rm H}\bm{\Pi}_k\bm{F}_k))$ being a monotonically decreasing function with respect to ${\boldsymbol\lambda}(\bm{F}_k^{\rm H}\bm{\Pi}_k\bm{F}_k)$ for $k=1,\cdots,K$. The optimal $\bm{F}_k$ is a Pareto optimal solution of the following vector optimization subproblem
\begin{align}
 \textbf{Opt.\,1.3:} \ \ \max\limits_{\bm{F}_k}  \ \ &{\boldsymbol\lambda}(\bm{F}_k^{\rm H}\bm{\Pi}_k\bm{F}_k) , \nonumber \\
 {\rm{s.t.}} \ \ \ \  &\psi_{k,i}(\bm{F}_k)\le 0, \ \ 1\le i\le I_k,
\end{align}which is equivalent to the following matrix-monotonic optimization problem \cite{XingTSP201501,Matrixmonotonic}
\begin{align}
 \textbf{Opt.\,1.4:} \ \ \max\limits_{\bm{F}_k}  \ \ &\bm{F}_k^{\rm H}\bm{\Pi}_k\bm{F}_k, \nonumber \\
 {\rm{s.t.}} \ \ \ \  &\psi_{k,i}(\bm{F}_k)\le 0, \ \ 1\le i\le I_k.
\end{align}where $\bm{\Pi}_k$ is independent of $\bm{F}_k$. Then, based on the results of Part I~\cite{Matrixmonotonic}, the optimal structure of $\bm{F}_k$ can be derived. Based on the optimal structures, the optimization problem can be
substantially simplified.  To elaborate a little further, given the
optimal structures, the optimization problem \textbf{Opt.\,1.2}
associated with multiple matrix variables can be efficiently solved in
an iterative manner. It is worth noting that given the
  optimal structures, an iterative algorithm is still needed to solve
  \textbf{Opt.\,1.2} and in most cases the iterative algorithms used
  are iterative water-filling algorithms
  \cite{General_Waterfilling,Palomar_Waterfilling}. Suffice to say
  that the convergence of this kind of algorithms can be guaranteed,
  but in a general case after convergence only covergence to the local
  optimum of the final solutions can be guaranteed.  Based on Part I
  \cite{Matrixmonotonic}, in the following the fundamental results for
  \textbf{Opt.\,1.4} are given, which constitute the basis for the
  following sections.

\underline{\textbf{Shaping Constraint}}
 For the shaping constraint, \textbf{Opt.\,1.4} becomes
 the following optimization problem \cite{XingTSP201502}
\begin{align}\label{eq30}
 \textbf{Opt.\,1.5:} \ \  {\max}_{\bm{F}_k}\ \ & \bm{F}_k^{\rm H}\bm{\Pi}_k\bm{F}_k \nonumber \\
  {\rm s.t.} \ &\bm{F}_k\bm{F}_k^{\rm H}\preceq \bm{R}_{\rm {s}_k}.
\end{align}
 The following lemma reveals the optimal structure of $\bm{F}_k$ for \textbf{Opt.\,1.5}
 with the shaping constraint.

\begin{lemma}\label{L1}
 When the rank of $\bm{R}_{\rm {s}_k}$ is not higher than the number of columns and the number
 of rows in $\bm{F}_k$, the optimal solution $\bm{F}_{\rm {opt},k}$ of \textbf{Opt.\,1.5} is a
 square root of $\bm{R}_{\rm {s}_k}$, i.e., $\bm{F}_{\rm {opt},k}\bm{F}_{\rm {opt},k}^{\rm H}=
 \bm{R}_{\rm {s}_k}$.
\end{lemma}

 \underline{\textbf{Joint Power Constraint}}
 Under the joint power constraint, \textbf{Opt.\,1.4} can be
 rewritten as
\begin{align}\label{eq31}
 \textbf{Opt.\,1.6:} \ \ {\max}_{\bm{F}_k} \ & \bm{F}_k^{\rm H}\bm{\Pi}_k\bm{F}_k \nonumber \\
   {\rm s.t.} & {\rm Tr}\big(\bm{F}_k\bm{F}_k^{\rm H}\big) \hspace{-1mm}\le\hspace{-1mm} P_k , \bm{F}_k\bm{F}_k^{\rm H}\preceq \tau_k \bm{I} .
\end{align}
 The Pareto optimal solution $\bm{F}_{\rm {opt},k}$ for \textbf{Opt.\,1.6} is given in Lemma~\ref{L2}.

\begin{lemma}\label{L2}
 For \textbf{Opt.\,1.6} with the joint power constraint, the Pareto optimal solutions
 satisfy the following structure
\begin{align}\label{eq32}
 \bm{F}_{\rm {opt},k} =& \bm{U}_{\bm{\Pi}_k}\bm{\Lambda}_{\bm{F}_k}\bm{U}_{\text{Arb},k}^{\rm H} ,
\end{align}
 where the unitary matrix $\bm{U}_{\bm{\Pi}_k}$ is specified by the EVD
\begin{align}\label{eq33}
 \bm{\Pi}_k =& \bm{U}_{\bm{\Pi}_k}\bm{\Lambda}_{\bm{\Pi}_k}\bm{U}_{\bm{\Pi}_k}^{\rm H} \
  \text{with} \ \bm{\Lambda}_{\bm{\Pi}_k} \searrow ,
\end{align}
 every diagonal element of the rectangular diagonal matrix $\bm{\Lambda}_{\bm{F}_k}$ is
 smaller than $\sqrt{\tau_k}$, and $\bm{U}_{\text{Arb},k}$ is an arbitrary unitary matrix
 having the appropriate dimension.
\end{lemma}

\underline{\textbf{Multiple Weighted Power Constraints}}
 Under the multiple weighted power constraints, \textbf{Opt.\,1.4} becomes
\begin{align}\label{eq34}
 \textbf{Opt.\,1.7:} \max\limits_{\bm{F}_k} \ \ & \bm{F}_k^{\rm H}\bm{\Pi}_k\bm{F}_k\nonumber \\
 {\rm s.t.} \ \ &{\rm Tr}\big(\bm{\Omega}_{k,i}\bm{F}_k\bm{F}_k^{\rm H}\big)\hspace{-1mm} \le \hspace{-1mm}P_{k,i} , 1\le i \le I_k .
\end{align}
 Note that the weighted power constraints include both the sum power constraint
 and per-antenna power constraints as its special cases. The Pareto optimal solution $\bm{F}_{\rm {opt},k}$ for \textbf{Opt.\,1.7} is given in Lemma~\ref{L3}.

\begin{lemma}\label{L3}
 The Pareto optimal solutions of \textbf{Opt.\,1.6} satisfy the following structure
\begin{align}\label{eq36}
 \bm{F}_{\rm {opt},k} =& \bm{\Omega}_k^{-\frac{1}{2}}\bm{U}_{\widetilde{\bm{\Pi}}_k}
  \bm{\Lambda}_{\widetilde{\bm{F}}_k}\bm{U}_{\rm {Arb},k}^{\rm H} ,
\end{align}
 where $\bm{U}_{\rm{ Arb},k}$ is an arbitrary unitary matrix of appropriate dimension,
 $\bm{\Omega}_k=\sum_{i=1}^{I_k}\alpha_{k,i}\bm{\Omega}_{k,i}$, the nonnegative scalars
 $\alpha_{k,i}$ are the weighting factors that ensure that the constraints ${\rm Tr}\big(\bm{\Omega}_{k,i}
 \bm{F}_k\bm{F}_k^{\rm H}\big)\le P_{k,i}$ hold and they can be computed by classic subgradient methods,
 while the unitary matrix $\bm{U}_{\widetilde{\bm{\Pi}}_k}$ is specified by the EVD
\begin{align}\label{eq37}
 \bm{\Omega}_k^{-\frac{1}{2}}\bm{\Pi}_k\bm{\Omega}_k^{-\frac{1}{2}} =& \bm{U}_{\widetilde{\bm{\Pi}}_k}
  \bm{\Lambda}_{\widetilde{\bm{\Pi}}_k}\bm{U}_{\widetilde{\bm{\Pi}}_k}^{\rm H} \ \text{with} \
  \bm{\Lambda}_{\widetilde{\bm{\Pi}}_k} \searrow .
\end{align}
\end{lemma}

In this paper, we focus our attention on the optimization
  problems of multiple complex matrix variates. In order to overcome
  the difficulties arising from the coupling relationships among the
  multiple matrix variates, the right unitarily invariant property of
  the constraints in \textbf{Opt.\,1.1} is exploited to introduce a
  series of auxiliary unitary matrices. Each auxiliary unitary matrix
  aligns its corresponding matrix variable to achieve extreme
  objective values. As a result, the optimal solutions of the matrix
  variables are Pareto optimal solutions of a series of single-variate
  matrix monotonic optimization problems. Then the optimal structure
  of each matrix variable can be derived, based on which the original
  optimization problem can be solved efficiently in an iterative
  manner. In the following, three specific optimization problems will
  be investigated, namely transceiver optimization for the multi-user
  MIMO (MU-MIMO) uplink, signal compression for distributed sensor
  networks and transceiver optimizations for multi-hop
  amplify-and-forward (AF) MIMO relaying networks. Generally speaking,
  an auxiliary unitary matrix aligns its lefthand side and righthand
  side with its corresponding matrix variables.  The three examples
  are specifically chosen for characterizing the effects of the matrix
  variates on the auxiliary unitary matrices. Specifically, in the
  transceiver optimization of the MU-MIMO uplink, when optimizing the
  $k$th matrix variate, the other matrix variates only affect
  the righthand side of the corresponding unitary matrix. As for
  signal compression in distributed sensor networks, when optimizing
  the $k$th matrix variate, the effects of other matrix
  variates are only on the lefthand side of the corresponding unitary
  matrix. Finally, as for transceiver optimizations in AF MIMO
  relaying networks, when optimizing the $k$th matrix variate,
  the other matrix variates affect both sides of the corresponding
  unitary matrix.

\section{MU-MIMO Uplink Communications}\label{S6}


 The first application scenario for the matrix monotonic optimization theory is found in MU
 MIMO uplink communications. In the MU MIMO uplink system of Fig.~\ref{Fig3},
 $K$ multi-antenna aided mobile users communicate with a multi-antenna assisted base
 station (BS) \cite{GoldsmithTextbook,DavidTSETextbook,NeworkInformationTheory,Tse2004}. The BS recovers the signals transmitted from all the $K$ mobile
 terminals. The sum rate maximization problem associated with this MU-MIMO
 uplink can be formulated as follows \cite{WYu2007,GoldsmithTextbook,DavidTSETextbook,NeworkInformationTheory}
\begin{align}
\label{Opt_MU_MIMO}
\hspace*{-2mm}\begin{array}{lcl}
  \textbf{Opt.\,2.1:}\!\!& \!\!\!\! \min\limits_{\{\bm{P}_k\}_{k=1}^K}\!\!\!\! &\!\!\!\!  -\log\bigg| \bm{R}_{\rm n}
  + \sum\limits_{k=1}^K \bm{H}_k \bm{X}_k{\bm{W}}_k\bm{X}_k^{\rm H} \bm{H}_k^{\rm H}\bigg| , \\
 \!\!\!\! &\!\!\!\! \rm{s.t.}\!\!\!\! &\!\!\!\! \psi_{k,i}(\bm{X}_k)\le 0, 1\le i\le I_k, 1\le k\le K ,
\end{array} \!\!
\end{align}where $\bm{H}_k$ is the MIMO channel matrix between the $k$th user and the BS,
 $\bm{X}_k$ is the precoding matrix at the $k$th user, and $\bm{R}_{\rm n}$ is
 the additive noise's covariance matrix at the BS. For the $k$th user, the positive definite matrix $\bm{W}_k$ is the corresponding weighting matrix.  Different from the work in
 \cite{WYu2007}, the power constraints considered in our work are more general
 than the per-antenna power constraints in \cite{WYu2007}. Similar to \textbf{Opt.\,1.2}, defining $\bm{X}_k=\bm{F}_k{\bm{Q}}_{\bm{X}_k}$, the optimization problem (\ref{Opt_MU_MIMO}) is equivalent to
 \begin{align}\label{eq87}
\hspace*{-2mm}\begin{array}{lcl}
 \textbf{Opt.\,2.2:}\!\!\!\! &\!\!\!\! \min\limits_{\{\bm{F}_k\}_{k=1}^K}\!\!\!\! &\!\!\!\!  -\log\bigg| \bm{R}_{\rm n}
  + \sum\limits_{k=1}^K \bm{H}_k \bm{F}_k{\bm{Q}}_{\bm{X}_k}{\bm{W}}_k{\bm{Q}}_{\bm{X}_k}^{\rm{H}}\bm{F}_k^{\rm H} \bm{H}_k^{\rm H}\bigg| , \\
 \!\!\!\! &\!\!\!\! \rm{s.t.}\!\!\!\! &\!\!\!\! \psi_{k,i}(\bm{F}_k)\le 0, 1\le i\le I_k, 1\le k\le K ,
\end{array} \!\!
\end{align}
 The objective function
 of \textbf{Opt.\,2.2} satisfies the following property, which can be exploited to
 optimize the multiple\begin{figure}[!h]
\vspace*{-4mm}
\begin{center}
\includegraphics[width=0.34\textwidth]{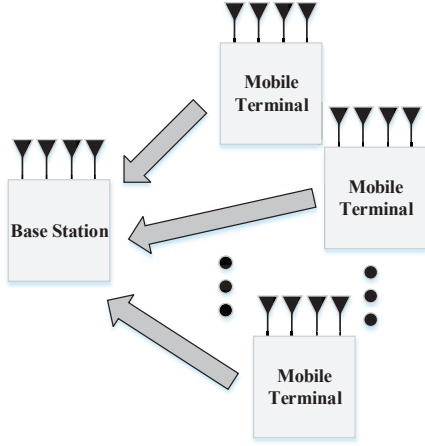}
\end{center}
\vspace*{-6mm}
\caption{The uplink of MU-MIMO communications.}
\label{Fig3}
\end{figure}matrix variables
\begin{align}\label{eq88}
 & \log \bigg| \bm{R}_{\rm n} + \sum\limits_{k=1}^K \bm{H}_k  \bm{F}_k{\bm{Q}}_{\bm{X}_k} \bm{W}_k {\bm{Q}}_{\bm{X}_k}^{\rm{H}}\bm{F}_k^{\rm H}
  \bm{H}_k^{\rm H}\bigg| \nonumber \\
 & = \log \bigg| \bm{I} + \bm{H}_k\bm{F}_k{\bm{Q}}_{\bm{X}_k}\bm{W}_k{\bm{Q}}_{\bm{X}_k}^{\rm{H}}\bm{F}_k^{\rm H}\bm{H}_k^{\rm H}
 \nonumber \\
 & \ \ \ \ \ \ \ \ \ \ \ \ \ \ \  \times\bigg( \hspace*{-1mm}  \bm{R}_{\rm n} +  \hspace*{-1mm} \sum\limits_{j\neq k} \hspace*{-1mm}  \bm{H}_j\bm{F}_j{\bm{Q}}_{\bm{X}_k} \bm{W}_j {\bm{Q}}_{\bm{X}_k}^{\rm{H}}\bm{F}_j^{\rm H}
  \bm{H}_j^{\rm H}\bigg)^{-1}    \bigg| \nonumber \\
 &  + \log\bigg| \bm{R}_{\rm n} + \sum\limits_{j\neq k}  \hspace*{-1mm} \bm{H}_j\bm{F}_j {\bm{Q}}_{\bm{X}_j}\bm{W}_j{\bm{Q}}_{\bm{X}_j}^{\rm{H}}
  \bm{F}_j^{\rm H}\bm{H}_j^{\rm H}\bigg| \nonumber \\
 & = \log\bigg| \bm{I} + \bm{W}_k{\bm{Q}}_{\bm{X}_k}^{\rm{H}} \bm{F}_k^{\rm H}\bm{H}_k^{\rm H}\bm{K}_{{\rm n}_k}^{-1}
  \bm{H}_k\bm{F}_k {\bm{Q}}_{\bm{X}_k}\bigg | + \log\big| \bm{K}_{{\rm n}_k} \big| ,
\end{align}
 where we have
\begin{align}\label{eq89}
 \bm{K}_{{\rm n}_k} =& \bm{R}_{\rm n} +\sum\nolimits_{j\neq k}
  \bm{H}_j\bm{F}_j{\bm{Q}}_{\bm{X}_j}\bm{W}_j{\bm{Q}}_{\bm{X}_j}^{\rm{H}}\bm{F}_j^{\rm H}\bm{H}_j^{\rm H}.
\end{align}

Therefore, based on (\ref{eq88}) for the $k$th matrix variate $\bm{F}_k$ \textbf{Opt.\,2.2} can be written in the following equivalent formula
\begin{align}
\hspace*{-2mm}\begin{array}{lcl}
  \textbf{Opt.\,2.3:} & \min\limits_{\bm{F}_k} & \hspace*{-2mm}-\log\bigg| \bm{I} + \bm{W}_k{\bm{Q}}_{\bm{X}_k}^{\rm{H}} \bm{F}_k^{\rm H}\bm{H}_k^{\rm H}\bm{K}_{{\rm n}_k}^{-1}
  \bm{H}_k\bm{F}_k {\bm{Q}}_{\bm{X}_k} \bigg | , \\
 & \rm{s.t.} & \bm{K}_{{\rm n}_k}  \hspace*{-1mm}=  \hspace*{-1mm} \bm{R}_{\rm n} \hspace*{-1mm} +  \hspace*{-1mm}\sum\limits_{j\neq k}  \hspace*{-1mm}\bm{H}_j\bm{F}_j{\bm{Q}}_{\bm{X}_j} \bm{W}_j{\bm{Q}}_{\bm{X}_j}^{\rm{H}}
  \bm{F}_j^{\rm H} \bm{H}_j^{\rm H} , \\
 & & \psi_{k,i}(\bm{F}_k)\le 0 , i=1,\cdots ,I_k .
\end{array}
\end{align} The matrix $\bm{F}_k^{\rm H}\bm{H}_k^{\rm H}\bm{K}_{{\rm n}_k}^{-1}\bm{H}_k\bm{F}_k$
 can be interpreted as the matrix version SNR for the $k$th user \cite{XingTSP201601}. Based on \textbf{Matrix Inequality 4} in Part I \cite{Matrixmonotonic}, we have
 \begin{align}
& \log\bigg| \bm{I} + \bm{W}_k{\bm{Q}}_{\bm{X}_k}^{\rm{H}} \bm{F}_k^{\rm H}\bm{H}_k^{\rm H}\bm{K}_{{\rm n}_k}^{-1}
  \bm{H}_k\bm{F}_k{\bm{Q}}_{\bm{X}_k} \bigg | \nonumber \\
&\le \sum_i \log\left(1+\lambda_i(\bm{W})\lambda_i( \bm{F}_k^{\rm H}\bm{H}_k^{\rm H}\bm{K}_{{\rm n}_k}^{-1}
  \bm{H}_k\bm{F}_k )\right).
 \end{align} The equality holds when the unitary matrix ${\bm{Q}}_{\bm{X}_k}$ equals
 \begin{align}
 \label{equ_Q_1}
 {\bm{Q}}_{{\rm{opt}},\bm{X}_k}={\bm{U}}_{{\rm{SNR}},k}{\bm{U}}_{\bm{W}}^{\rm{H}}
\end{align} where the unitary matrices ${\bm{U}}_{{\rm{SNR}},k}$ and ${\bm{U}}_{\bm{W}_k}$ are defined based on the following EVDs
\begin{align}
&\bm{F}_k^{\rm H}\bm{H}_k^{\rm H}\bm{K}_{{\rm n}_k}^{-1}
  \bm{H}_k\bm{F}_k ={\bm{U}}_{{\rm{SNR}},k}{\bm{\Lambda}}_{{\rm{SNR}},k}{\bm{U}}_{{\rm{SNR}},k}^{\rm{H}}
  \text{ with }  \bm{\Lambda}_{{\rm{SNR}},k} \searrow
  \nonumber \\
 & \bm{W}_k = \bm{U}_{\bm{W}_k}
  \bm{\Lambda}_{\bm{W}_k} \bm{U}_{\bm{W}_k}^{\rm H}
  \text{ with }  \bm{\Lambda}_{\bm{W}_k} \searrow. \label{Unitary_Matrix_SNR}
\end{align} From the multi-objective optimization viewpoint, the optimal solutions of
 \textbf{Opt.\,2.3} belong to the Pareto optimal solution sets of the following
 optimization problems for $1\le k\le K$
\begin{align}
\hspace*{-2mm}\begin{array}{lcl}
  \textbf{Opt.\,2.4:} & \max\limits_{\bm{F}_k} & {\boldsymbol \lambda}\left(\bm{F}_k^{\rm H}\bm{H}_k^{\rm H}\bm{K}_{{\rm n}_k}^{-1}
  \bm{H}_k\bm{F}_k\right) , \\
 & \rm{s.t.} & \bm{K}_{{\rm n}_k}  \hspace*{-1mm}=  \hspace*{-1mm} \bm{R}_{\rm n} \hspace*{-1mm} +  \hspace*{-1mm}\sum\limits_{j\neq k}  \hspace*{-1mm}\bm{H}_j\bm{F}_j{\bm{Q}}_{\bm{X}_j} \bm{W}_j{\bm{Q}}_{\bm{X}_j}^{\rm{H}}
  \bm{F}_j^{\rm H} \bm{H}_j^{\rm H} , \\
 & & \psi_{k,i}(\bm{F}_k)\le 0 , i=1,\cdots ,I_k .
\end{array}
\end{align} which is equivalent to
\begin{align}\label{eq90}
\begin{array}{lcl}
 \textbf{Opt.\,2.5:} & \max\limits_{\bm{F}_k} & \bm{F}_k^{\rm H}\bm{H}_k^{\rm H}\bm{K}_{{\rm n}_k}^{-1}
  \bm{H}_k\bm{F}_k , \\
 & \rm{s.t.} & \bm{K}_{{\rm n}_k}  \hspace*{-1mm}=  \hspace*{-1mm} \bm{R}_{\rm n} \hspace*{-1mm} +  \hspace*{-1mm}\sum\limits_{j\neq k}  \hspace*{-1mm}\bm{H}_j\bm{F}_j{\bm{Q}}_{\bm{X}_j} \bm{W}_j{\bm{Q}}_{\bm{X}_j}^{\rm{H}}
  \bm{F}_j^{\rm H} \bm{H}_j^{\rm H} , \\
 & & \psi_{k,i}(\bm{F}_k)\le 0 , i=1,\cdots ,I_k .
\end{array}
\end{align}
 It can be seen that by using alternating optimization algorithm, the multiple-matrix-variate
optimization of \textbf{Opt.\,2.3} is transferred into the multiple
 single-matrix-variate matrix-monotonic optimization of \textbf{Opt.\,2.5}. Based on \textbf{Opt.\,2.5}, the optimal structure of $\bm{F}_k$ can be derived, and then the original optimization problem \textbf{Opt.\,2.2} can be solved in an iterative manner. It is worth noting that in most cases, for the alternating optimization algorithm, the final solutions are suboptimal. The alternating optimization algorithm stops when
the performance improvement is smaller than a predefined threshold or the iteration number reaches the predefined maximum value.  The convergence can be guaranteed when the subproblems are solved with global optimality.

{\emph{1)}~{\rm\bf Shaping Constraint}}: We have $I_k=1$ and
\begin{align}\label{eq91}
 \psi_{k,1}(\bm{F}_k) =& \bm{F}_k\bm{F}_k^{\rm H} - \bm{R}_{{\rm s}_k} .
\end{align}
 Based on Lemma~1 in Section~\ref{S2}, we readily conclude that for $1\le k\le K$, when the rank of
 $\bm{R}_{{\rm s}_k}$ is not higher than the number of columns and the number of rows
 in $\bm{F}_k$, the optimal solution $\bm{F}_{{\rm opt},k}$ of \textbf{Opt.\,2.3} is a
 square root of $\bm{R}_{{\rm s}_k}$.

{\emph{2)}~{\rm\bf Joint Power Constraint}:
 We have $I_k=2$ and
\begin{align}\label{eq92}
\begin{array}{l}
 \psi_{k,1}(\bm{F}_k) = {\rm Tr}\big(\bm{F}_k\bm{F}_k^{\rm H}\big) - P_k , \\
 \psi_{k,2}(\bm{F}_k) = \bm{F}_k\bm{F}_k^{\rm H} - \tau_k \bm{I} .
\end{array}
\end{align}
 Based on Lemma~2 in Section~\ref{S2}, we readily conclude that for $1\le k\le K$, the optimal solution
 $\bm{F}_{{\rm opt},k}$ of \textbf{Opt.\,2.3} satisfies the following structure
\begin{align}\label{eq93}
 \bm{F}_{{\rm opt},k} =& \bm{V}_{\widetilde{\bm{H}}_k} \bm{\Lambda}_{\bm{F}_k} \bm{U}_{{\rm{Arb}},k}^{\rm H} ,
\end{align}
 where the unitary matrix $\bm{V}_{\widetilde{\bm{H}}_k}$ is defined based on the SVD
\begin{align}\label{eq94}
 \bm{K}_{{\rm n}_k}^{-\frac{1}{2}}\bm{H}_k =& \bm{U}_{\widetilde{\bm{H}}_k}
  \bm{\Lambda}_{\widetilde{\bm{H}}_k} \bm{V}_{\widetilde{\bm{H}}_k}^{\rm H}
  \text{ with }  \bm{\Lambda}_{\widetilde{\bm{H}}_k} \searrow.
\end{align}
 and every diagonal element of the rectangular diagonal matrix $\bm{\Lambda}_{\bm{F}_k}$ is
 smaller than $\sqrt{\tau_k}$. The diagonal matrix $\bm{\Lambda}_{\bm{F}_k}$ can be efficiently solved using a variant water-filling algorithm \cite{FFGaoTSP,General_Waterfilling}.

{\emph{3)}~{\rm\bf Multiple Weighted Power Constraints}:
 In this case, we have
\begin{align}\label{eq95}
 \psi_{k,i}(\bm{F}_k) =& {\rm Tr}\big(\bm{\Omega}_{k,i} \bm{F}_k\bm{F}_k^{\rm H}\big) - P_{k,i} .
\end{align}\begin{figure}[!h]
\vspace*{-5mm}
\begin{center}
\includegraphics[width=0.4\textwidth]{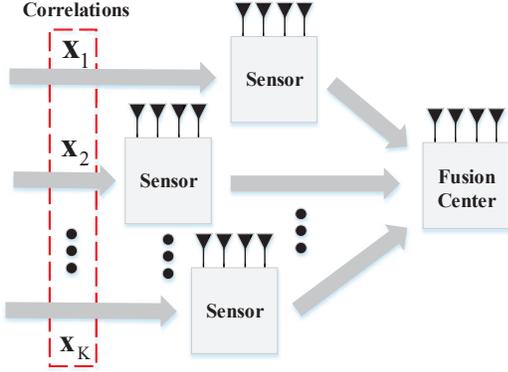}
\end{center}
\vspace*{-6mm}
\caption{Illustration of distribute sensor network.}
\label{Fig4}
\vspace*{-5mm}
\end{figure}Then
 based on Lemma~3 in Section~\ref{S2}, we conclude that for $1\le k\le K$, the optimal solution
 $\bm{F}_{{\rm opt},k}$ of \textbf{Opt.\,2.3}  satisfies the following structure
\begin{align}\label{eq96}
 \bm{F}_{{\rm opt},k} =& \bm{\Omega}_k^{-\frac{1}{2}}\bm{V}_{\bm{\mathcal{H}}_k}
  \bm{\Lambda}_{\widetilde{\bm{F}}_k} \bm{U}_{{\rm{Arb}},k}^{\rm H} ,
\end{align}where the unitary matrix $\bm{V}_{\bm{\mathcal{H}}_k}$ is defined by the following SVD
\begin{align}\label{eq97}
 \bm{K}_{{\rm n}_k}^{-\frac{1}{2}}\bm{H}_k\bm{\Omega}_k^{-\frac{1}{2}} =& \bm{U}_{\bm{\mathcal{H}}_k}
  \bm{\Lambda}_{\bm{\mathcal{H}}_k}\bm{V}_{\bm{\mathcal{H}}_k}^{\rm H} \text{ with }
  \bm{\Lambda}_{\bm{\mathcal{H}}_k} \searrow ,
\end{align}and the matrix $\bm{\Omega}_k$ is defined as
\begin{align}\label{eq98}
 \bm{\Omega}_k =& \sum\nolimits_{i=1}^{I_k} \alpha_{k,i} \bm{\Omega}_{k,i} .
\end{align} The diagonal matrix $ \bm{\Lambda}_{\widetilde{\bm{F}}_k}$ can be efficiently solved using water-filling algorithms \cite{Palomar_Waterfilling}.

\section{Signal Compression for Distributed Sensor Networks}\label{S7}

 In the distributed sensor network illustrated in Fig.~\ref{Fig4}, the $K$ sensors
 transmit their individual signals to the fusion center \cite{Fang2012,Fang2008,Fang2009,SensorBehbahani2012,SensorChawla2019,SensorGuo2015,SensorLI2007,SensorLiu2019,SensorNordio2015,SensorVenkategowda2018}. Specifically, the $k$th
 sensor transmits its signal $\bm{x}_k$ to the fusion center, when the channel between
 the $k$th sensor and the fusion center is $\bm{H}_k$. The fusion center recovers the
 transmitted signals $\bm{x}_k$ for $1\le k\le K$. In contrast to the scenario of
 MU-MIMO communications, there exist correlations among $\bm{x}_k$ \cite{JFang2013},
 and the correlation matrix is denoted by \setcounter{equation}{36}
\begin{align}\label{eq106}
 \bm{C}_{\bm{x}} =& \mathbb{E}\Big\{\big[\bm{x}_1^{\rm T} \cdots \bm{x}_K^{\rm T}\big]^{\rm T}
  \big[\bm{x}_1^{\rm T} \cdots \bm{x}_K^{\rm T}\big]^{*} \Big\} .
\end{align}
 Note that the correlations among the signals make the optimization approach of this
 application totally different from that of the MU-MIMO application.

To maximize the mutual information between the received signal at the data fusion center and the signal to estimate, the signal
 compression can be formulated as \textbf{Opt.\,3.1} \cite{JFang2013}, given as
 \begin{align}\label{eq107}
 \hspace*{-3mm} \textbf{Opt.\,3.1:} \min\limits_{\{\bm{X}_k\}_{k=1}^K} &  \hspace*{-2mm} -\log\Big| \bm{C}_{\bm{x}}^{-1} \nonumber \\
 & +
  {\rm diag}\Big\{ \big\{\bm{X}_k^{\rm H}\bm{H}_k^{\rm H}\bm{R}_{\rm{n}_k}^{-1}\bm{H}_k
  \bm{X}_k\big\}_{k=1}^K\Big\}\Big| , \nonumber \\
  {\rm s.t.} & \psi_{k,i}\big(\bm{X}_k\bm{R}_{\bm{x}_k}^{\frac{1}{2}}\big) \le 0, ~
  1\le i\le I_k, 1\le k\le K ,
\end{align} where $\bm{F}_k$ is the signal compression matrix at the $k$th sensor, $\bm{R}_{\bm{x}_k}$ is the covariance matrix of the signal $\bm{x}_k$ transmitted from
 the $k$th sensor, and $\bm{R}_{\rm{n}_k}$ is the covariance matrix of the additive
 noise $\bm{n}_k$ for the $k$th sensor signal received in its own time slot at the fusion center. Note
 that if all the sensors send signals at the same frequency, all the $\bm{R}_{\rm{n}_k}$
 are identical. If the sensors use different frequency bands, the noise covariance
 matrices $\bm{R}_{{\rm{n}}_k}$ are different.


 Note that in \cite{JFang2013}, only the simple sum power constraint is considered, while
 in our work the more general multiple weighted linear power constraints are taken into
 account. In other words, the result derived in this section for signal compression in
 distributed sensor networks is novel.

 For the general correlation matrix $\bm{C}_{\bm{x}}$, it is difficult to directly
 decouple the optimization problem. A natural choice is to take advantage of alternating
 optimization algorithms among $\bm{X}_k$ for $1\le k\le K$. To simplify the derivation,
 a permutation matrix $\bm{P}_k$ is first introduced, which reorders the block diagonal
 matrix ${\rm diag}\Big\{ \big\{\bm{X}_k^{\rm H}\bm{H}_k^{\rm H}\bm{R}_{\rm{n}_k}^{-1}\bm{H}_k
  \bm{X}_k\big\}_{k=1}^K\Big\}$ so that the following equality holds\setcounter{equation}{38}
\begin{align}\label{eq108}
 & \bm{P}_k{\rm diag}\Big\{ \big\{\bm{X}_k^{\rm H}\bm{H}_k^{\rm H}\bm{R}_{\rm{n}_k}^{-1}\bm{H}_k
  \bm{X}_k\big\}_{k=1}^K\Big\} \bm{P}_k^{\rm H} \nonumber \\
 & \hspace*{10mm}= \left[ \begin{array}{cc}
  \bm{X}_k^{\rm H}\bm{H}_k^{\rm H}\bm{R}_{\rm{n}_k}^{-1}\bm{H}_k \bm{X}_k & \bm{0} \\
  \bm{0} & \bm{\Xi}_k \end{array} \right] .
\end{align}The computation of $\bm{P}_k$ and the definition of $\bm{\Xi}_k$ are provided in Appendix~\ref{Appendix_Sensor_Precoding}. The permutation matrix $\bm{P}_k$ aims at moving the term $\bm{X}_k^{\rm H}\bm{H}_k^{\rm H}\bm{R}_{\rm{n}_k}^{-1}\bm{H}_k
  \bm{X}_k$ at the top of the block diagonal matrix.  The permutation
  matrix $\bm{P}_k$ is determined by the position of the term
  $\bm{X}_k^{\rm H}\bm{H}_k^{\rm H}\bm{R}_{\rm{n}_k}^{-1}\bm{H}_k
  \bm{X}_k$ in the block diagonal matrix ${\rm diag}\Big\{
  \big\{\bm{X}_k^{\rm H}\bm{H}_k^{\rm H}\bm{R}_{\rm{n}_k}^{-1}\bm{H}_k
  \bm{X}_k\big\}_{k=1}^K\Big\}$.  Note that a permutation matrix is
also a unitary matrix. By further exploiting the properties of matrix
determinants, \textbf{Opt.\,3.1} becomes equivalent to
\textbf{Opt.\,3.2} of (\ref{eq109}).
\begin{align}\label{eq109}
 \hspace*{-2mm}\textbf{Opt.\,3.2:}  \min\limits_{\{\bm{F}_k\}_{k=1}^K}  \hspace*{-3mm}&-\log\Big| \bm{P}_k\bm{C}_{\bm{x}}^{-1}\bm{P}_k^{\rm H}\hspace*{-1mm} \nonumber \\
 &+\hspace*{-1mm}
  \bm{P}_k{\rm diag}\Big\{ \big\{\bm{X}_k^{\rm H}\bm{H}_k^{\rm H}\bm{R}_{\rm{n}_k}^{-1}\bm{H}_k
  \bm{X}_k\big\}_{k=1}^K\Big\}\bm{P}_k^{\rm H}\Big| , \nonumber \\
 {\rm s.t.} & \psi_{k,i}\big(\bm{X}_k\bm{R}_{\bm{x}_k}^{\frac{1}{2}}\big) \le 0, ~
  1\le i\le I_k, 1\le k\le K .
\end{align}

 In order to simplify \textbf{Opt.\,3.2}, we divide $\bm{P}_k\bm{C}_{\bm{x}}^{-1}\bm{P}_k^{\rm H}$
 into \setcounter{equation}{40}
\begin{align}\label{eq110}
 \bm{P}_k\bm{C}_{\bm{x}}^{-1}\bm{P}_k^{\rm H} =& \left[ {\begin{array}{cc}
  \bm{P}_{1,1} & \bm{P}_{1,2} \\ \bm{P}_{2,1} & \bm{P}_{2,2}
\end{array}} \right] .
\end{align}
 Combining (\ref{eq108}) and (\ref{eq110}) leads to
\begin{align}\label{eq111}
 & \bm{P}_k\bm{C}_{\bm{x}}^{-1}\bm{P}_k^{\rm H} + \bm{P}_k{\rm diag}\Big\{ \big\{\bm{X}_k^{\rm H}
  \bm{H}_k^{\rm H}\bm{R}_{\rm{n}_k}^{-1}\bm{H}_k \bm{X}_k\big\}_{k=1}^K\Big\}\bm{P}_k^{\rm H} \nonumber \\
 & \hspace*{6mm}= \left[ \begin{array}{cc}
  \bm{P}_{1,1} + \bm{X}_k^{\rm H}\bm{H}_k^{\rm H}\bm{R}_{\rm{n}_k}^{-1}\bm{H}_k \bm{X}_k
  & \bm{P}_{1,2} \\ \bm{P}_{2,1} & \bm{P}_{2,2} + \bm{\Xi}_k
\end{array} \right] \! .\!
\end{align}
 Further exploiting the fundamental properties of matrix determinants
 \cite{JFang2013,Horn1990}, we have the following equality
\begin{align}\label{eq112}
 & \left|\left[ \begin{array}{cc}
  \bm{P}_{1,1} + \bm{X}_k^{\rm H}\bm{H}_k^{\rm H}\bm{R}_{\rm{n}_k}^{-1}\bm{H}_k \bm{X}_k
  & \bm{P}_{1,2} \\ \bm{P}_{2,1} & \bm{P}_{2,2} + \bm{\Xi}_k
  \end{array} \right]\right| \nonumber \\
 & \hspace*{10mm}= \big|\bm{P}_{2,2} + \bm{\Xi}_k\big| \big|\bm{X}_k^{\rm H}\bm{H}_k^{\rm H}\bm{R}_{\rm{n}_k}^{-1}
  \bm{H}_k\bm{X}_k + \bm{\Phi}_k \big|,
\end{align}
 where
\begin{align}\label{eq113}
 \bm{\Phi}_k =& \bm{P}_{1,1} - \bm{P}_{1,2}(\bm{P}_{2,2} + \bm{\Xi}_k)^{-1}\bm{P}_{2,1} .
\end{align}
 Based on (\ref{eq112}), the alternating optimization of $\bm{F}_k$
 for $1\le k\le K$ can be performed. Specifically, the optimization
 problem \textbf{Opt.\,3.2} is transferred into: for $1\le k\le K$,
\begin{align}\label{eq114}
\hspace*{-2mm}\begin{array}{lcl}
 \textbf{Opt.\,3.3:} & \min\limits_{\bm{X}_k} & -\log\big|\bm{\Phi}_k + \bm{X}_k^{\rm H}
  \bm{H}_k^{\rm H}\bm{R}_{\bm{n}_k}^{-1}\bm{H}_k\bm{X}_k\big| , \\
 & {\rm{s.t.}} & \psi_{k,i}\big(\bm{X}_k\bm{R}_{\bm{x}_k}^{\frac{1}{2}}\big) \le 0 ,
  1\le i\le I_k .
\end{array}
\end{align}
 It can be seen that by exploiting its block diagonal structure, the multiple-matrix-variate
 matrix-monotonic optimization of \textbf{Opt.\,3.1} is transferred into several
 single-matrix-variate matrix-monotonic optimization problems in the form of \textbf{Opt.\,3.3}.

 For $1\le k\le K$, by introducing the auxiliary variable
\begin{align}\label{eq115}
 {\bm{F}}_k{\bm{Q}}_{\bm{X}_k} =& \bm{X}_k\bm{R}_{\bm{x}_k}^{\frac{1}{2}} ,
\end{align}
 the optimization problem \textbf{Opt.\,3.3} is transferred into:
\begin{align}\label{eq116}
\hspace*{-4mm}
 \textbf{Opt.\,3.4:}  \ \min\limits_{{\bm{F}}_k} & -\log\big| \bm{R}_{\bm{x}_k}^{-\frac{1}{2}}
  \bm{\Phi}_k\bm{R}_{\bm{x}_k}^{-\frac{1}{2}}  \nonumber \\
  &+
  \hspace*{-1mm}{\bm{Q}}_{\bm{X}_k}^{\rm{H}}{\bm{F}}_k^{\rm H}
  \bm{H}_k^{\rm H}\bm{R}_{\bm{n}_k}^{-1}\bm{H}_k{\bm{F}}_k{\bm{Q}}_{\bm{X}_k}\big| , \nonumber \\
{\rm{s.t.}}\!\! &\!\! \psi_{k,i}\big({\bm{F}}_k\big) \le 0 , 1\le i\le I_k .
\end{align}
Based on \textbf{Matrix Inequality 3} in Part I  \cite{Matrixmonotonic}, we have
\begin{align}
&\log\big| \bm{R}_{\bm{x}_k}^{-\frac{1}{2}}
  \bm{\Phi}_k\bm{R}_{\bm{x}_k}^{-\frac{1}{2}}\hspace*{-1mm} + \hspace*{-1mm}{\bm{Q}}_{\bm{X}_k}^{\rm{H}}{\bm{F}}_k^{\rm H}
  \bm{H}_k^{\rm H}\bm{R}_{\bm{n}_k}^{-1}\bm{H}_k{\bm{F}}_k{\bm{Q}}_{\bm{X}_k}\big| \nonumber \\
&\le \sum_j{\rm{log}}|\lambda_{N-j+1}(\bm{R}_{\bm{x}_k}^{-\frac{1}{2}}
  \bm{\Phi}_k\bm{R}_{\bm{x}_k}^{-\frac{1}{2}}) +\lambda_j({\bm{F}}_k^{\rm H}
  \bm{H}_k^{\rm H}\bm{R}_{\bm{n}_k}^{-1}\bm{H}_k{\bm{F}}_k) |
\end{align} based on which the optimal unitary matrix ${\bm{Q}}_{\bm{X}_k}$ equals \cite{XingTSP201501}
 \begin{align}
 {\bm{Q}}_{{\rm{opt}},\bm{X}_k}={\bm{U}}_{{\rm{SNR}},k}\bar{\bm{U}}_{\bm{\Phi}_k\bm{R}_k}^{\rm{H}}
\end{align} where the unitary matrices ${\bm{U}}_{{\rm{SNR}},k}$ and $\bar{\bm{U}}_{\bm{\Phi}_k\bm{R}_k}^{\rm H}$ are defined by the following SVD and EVD,
\begin{align}
&\bm{F}_k^{\rm H}\bm{H}_k^{\rm H}\bm{K}_{{\rm n}_k}^{-1}
  \bm{H}_k\bm{F}_k ={\bm{U}}_{{\rm{SNR}},k}{\bm{\Lambda}}_{{\rm{SNR}},k}{\bm{U}}_{{\rm{SNR}},k}^{\rm{H}}
  \text{ with }  \bm{\Lambda}_{{\rm{SNR}},k} \searrow
  \nonumber \\
 & \bm{R}_{\bm{x}_k}^{-\frac{1}{2}}\bm{\Phi}_k \bm{R}_{\bm{x}_k}^{-\frac{1}{2}}\! =\!
  \bar{\bm{U}}_{\bm{\Phi}_k\bm{R}_k} \bar{\bm{\Lambda}}_{\bm{\Phi}_k\bm{R}_k}
  \bar{\bm{U}}_{\bm{\Phi}_k\bm{R}_k}^{\rm H} \! \text{ with } \bar{\bm{\Lambda}}_{\bm{\Phi}_k\bm{R}_k}\! \nearrow \!.
\end{align}

From the multi-objective optimization viewpoint, the optimal solutions of
 \textbf{Opt.\,3.4} belong to the Pareto optimal solution sets of the following
 optimization problems for $1\le k\le K$ \cite{XingTSP201501}
\begin{align}
\hspace*{-2mm}\begin{array}{lcl}
 \textbf{Opt.\,3.5:}   & \max\limits_{{\bm{F}}_k} & {\boldsymbol \lambda}\left({\bm{F}}_k^{\rm H}
  \bm{H}_k^{\rm H}\bm{R}_{\rm{n}_k}^{-1}\bm{H}_k{\bm{F}}_k\right) , \\
 & {\rm{s.t.}} & \psi_{k,i}\big({\bm{F}}_k\big) \le 0 , 1\le i\le I_k .
\end{array}
\end{align} which is equivalent to the following matrix-monotonic optimization problem:
\begin{align}\label{eq117}
\hspace*{-2mm}\begin{array}{lcl}
 \textbf{Opt.\,3.6:}  & \max\limits_{{\bm{F}}_k} & {\bm{F}}_k^{\rm H}
  \bm{H}_k^{\rm H}\bm{R}_{\rm{n}_k}^{-1}\bm{H}_k{\bm{F}}_k , \\
 & {\rm{s.t.}} & \psi_{k,i}\big({\bm{F}}_k\big) \le 0 , 1\le i\le I_k .
\end{array}
\end{align}
 Based on the fundamental results of the previous sections derived for matrix-monotonic
 optimization, we have the following results.  Clearly, the optimal $ \bm{X}_{k}$ equals
 \begin{align}
 \bm{X}_{{\rm opt},k}=
 {\bm{F}}_{{\rm opt},k}{\bm{Q}}_{{\rm opt},\bm{X}_k} \bm{R}_{\bm{x}_k}^{-\frac{1}{2}}.
 \end{align}


\begin{figure*}[bp!]\setcounter{equation}{74}
\vspace*{-4mm}
\hrule
\begin{align}\label{Optimziation_Original} 
\begin{array}{lcl}
 \textbf{Opt\,4.1:} & \min\limits_{\{\bm{F}_k\}_{k=1}^{K},\{\bm{Q}_{\bm{X}_k}\}_{k=1}^{K},\bm{C}}
  & f\big(\bm{\Phi}_{\rm MSE}\big(\{\bm{F}_k\}_{k=1}^{K},\{\bm{Q}_{\bm{X}_k}\}_{k=1}^{K},\bm{C}\big)\big), \\
 & {\rm{s.t.}} & \psi_{k,i}(\bm{F}_k)\le 0, ~ 1\le i\le I_k, 1\le k\le K, \\
 & & [\bm{C}]_{i,i}=1 , ~ [\bm{C}]_{i,j}=0  ~ {\text{for}} \ \ i<j , 1\le i\le N .
\end{array}
\end{align}
\vspace*{-1mm}
\end{figure*}

{\emph{1)}~{\rm\bf Shaping Constraint}}: We have $I_k=1$ and\setcounter{equation}{62}
\begin{align}\label{eq118}
 \psi_{k,1}\big({\bm{F}}_k\big) = {\bm{F}}_k{\bm{F}}_k^{\rm H} -
  \bm{R}_{{\rm s}_k} .
\end{align}\begin{figure}[!h]
	\centering
	\includegraphics[width =0.48\textwidth]{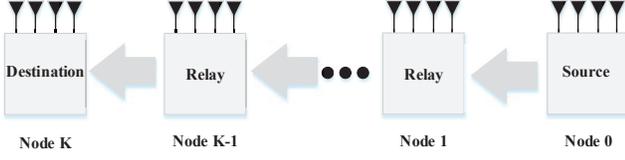}
	\caption{Multi-hop cooperative AF MIMO relaying network.}
	\label{Fig5}
\end{figure}Based on Lemma~\ref{L2} in Section~\ref{S2}, we have
 when the rank of $\bm{R}_{{\rm s}_k}$ is not higher than the number of columns and
 the number of rows in ${\bm{F}}_k$, the optimal solution ${\bm{F}}_{{\rm opt},k}$
 is a square root of $\bm{R}_{{\rm s}_k}$.

{\emph{2)}~{\rm\bf Joint Power Constraints}: We have
\begin{align}\label{eq119}
\begin{array}{l}
 \psi_{k,1}\big({\bm{F}}_k\big) = {\rm Tr}\big({\bm{F}}_k{\bm{F}}_k^{\rm H}\big) - P_k , \\
 \psi_{k,2}\big({\bm{F}}_k\big) = {\bm{F}}_k{\bm{F}}_k^{\rm H} - \tau_k \bm{I} .
\end{array}
\end{align}Based on Lemma~\ref{L2} in Section~\ref{S2},
 the Pareto optimal solutions $\bm{F}_{{\rm opt},k}$ satisfy the following structure
\begin{align}\label{eq120}
 \bm{F}_{{\rm opt},k} = \bm{V}_{\bm{H}_k}\bm{\Lambda}_{{\bm{F}}_k}
  \bm{U}_{{\text{Arb}},k}^{\rm H},
\end{align}where every diagonal element of the rectangular diagonal matrix $\bm{\Lambda}_{\breve{\bm{F}}_k}$
 is smaller than $\sqrt{\tau_k}$.  The diagonal matrix $\bm{\Lambda}_{\breve{\bm{F}}_k}$ can be efficiently solved using a variant water-filling algorithm \cite{General_Waterfilling,FFGaoTSP}.

{\emph{3)}~{\rm\bf Multiple Weighted Power Constraints}: We have
\begin{align}\label{eq121}
 \psi_{k,i}\big({\bm{F}}_k\big) = {\rm Tr}\big(\bm{\Omega}_{k,i}{\bm{F}}_k
  {\bm{F}}_k^{\rm H}\big) - P_{k,i} .
\end{align}Based on Lemma~\ref{L3} in Section~\ref{S2},
 the Pareto optimal solutions $\bm{F}_{{\rm opt},k}$ satisfy the following structure
\begin{align}\label{eq122}
 \bm{F}_{{\rm opt},k} = \bm{\Omega}_k^{-\frac{1}{2}} \breve{\bm{V}}_{\bm{\mathcal{H}}_k}
  \bm{\Lambda}_{\breve{\bm{F}}_k}{\bm{U}}_{{\rm{Arb}},k}^{\rm H},
\end{align}
 where $\bm{\Omega}_k$ is given by (\ref{eq98}), while $\breve{\bm{V}}_{\bm{\mathcal{H}}_k}$
 is  defined by the following SVD,
 respectively,
\begin{align}
 & \bm{R}_{\bm{n}_k}^{-\frac{1}{2}}\bm{H}_k\bm{\Omega}_k^{-\frac{1}{2}}\! =\!
  \breve{\bm{U}}_{\bm{\mathcal{H}}_k} \breve{\bm{\Lambda}}_{\bm{\mathcal{H}}_k}\breve{\bm{V}}_{\bm{\mathcal{H}}_k}^{\rm H}
  \text{ with } \breve{\bm{\Lambda}}_{\bm{\mathcal{H}}_k} \searrow. \label{eq123}
\end{align} The diagonal matrix $\bm{\Lambda}_{\breve{\bm{F}}_k}$ can be efficiently solved using water-filling algorithms \cite{Palomar_Waterfilling,General_Waterfilling}.

\begin{remark} The results of this paper can also be
    applied to more complex scenarios. For example, when the CSI
    between a sensor and its data fusion center is imperfect, $\bm{H}_k
    =\widehat{\bm{H}}_k + \bm{H}_{{\rm W},k}
    \bm{\Psi}_k^{\frac{1}{2}}$, where $\widehat{\bm{H}}_k$ and
    $\bm{H}_{{\rm W},k}\bm{\Psi}_k^{\frac{1}{2}}$ are the estimated
    CSI and the channel estimation error, respectively. The
    correlation matrix $\bm{\Psi}_k$ is a function of both the channel
    estimator and of the training sequence.  Based on the proposed framework,
    the optimal structures of the optimal solutions for the robust
    signal compression matrices at different sensors can also be readily
    derived.\end{remark}

\section{Multi-Hop AF MIMO Relaying Networks}\label{S8}

 Multi-hop relaying communication  is one of the most
 important enabling technologies for future flexible and high-throughput
 communications, such as machine-to-machine, device-to-device, vehicle-to-vehicle, internet of things
 or satellite communications \cite{XingTSP201502,Rong2009TWC}. The key idea behind multi-hop communications is to
 deploy multiple relays to realize the communications between the source node and
 destination node \cite{Rong2009TWC,Mo09}.
 Before presenting our third application of transceiver
 optimization for multi-hop communications, we first highlight the difference
 between our work presented in this section and the previous conclusions in
 \cite{XingTSP201502,XingTSP201601}.
\begin{itemize}
\item We consider a more general power constraint which includes both the
 per-antenna power constraint in \cite{XingTSP201601} and the shaping constraints
 in \cite{XingTSP201502} as its special cases.
\item The channel estimation errors are realistically taken into account
 in our work. By contrast, in \cite{XingTSP201601} the CSI is assumed to be perfectly
 known.
\end{itemize}
 To the best of our knowledge, the robust transceiver optimization for
 multi-hop communications even under the per-antenna power constraint is still
 the problem not yet fully solved in the existing literature. Therefore, the results presented
 in this section is novel and significant.

\begin{table*}[tp!]
\vspace*{-1mm}
\caption{The objective functions and the associated optimal first unitary matrices $\bm{Q}_1$
 for multi-hop cooperative AF relay networks.}
\vspace*{-4mm}
\begin{center}
\scalebox{1.2}{
\begin{tabular}{||l|l||l||}
\hline
 Index & Objective function & Optimal $\bm{Q}_{\bm{X}_1}$ \\ \hline\hline
 \textbf{Obj.\,1} & $\log\big|\bm{\Phi}_{\rm MSE}\big(\{\bm{F}_k\}_{k=1}^{K},\{\bm{Q}_{\bm{X}_k}\}_{k=1}^{K},
  \bm{C}=\bm{I}\big)|$ & $\bm{Q}_{{\rm opt},\bm{X}_1}=\bm{V}_{\bm{A}_1}\bm{U}_{\rm{Arb}}^{\rm H}$ \\ \hline
 \textbf{Obj.\,2} & ${\rm{Tr}}\big(\bm{W}\bm{\Phi}_{\rm MSE}\big(\{\bm{F}_k\}_{k=1}^{K},
  \{\bm{Q}_{\bm{X}_k}\}_{k=1}^{K},\bm{C}=\bm{I}\big)\big)$ & $\bm{Q}_{{\rm opt},\bm{X}_1}=\bm{V}_{\bm{A}_1}
  \bm{U}_{\bm{W}}^{\rm H}$ \\ \hline
 \textbf{Obj.\,3} & $f_{\text{A-Schur}}^{\text{Convex}}\big(\bm{d}\big[\bm{\Phi}_{\rm MSE}\big(\{\bm{F}_k\}_{k=1}^{K},
  \{\bm{Q}_{\bm{X}_k}\}_{k=1}^{K},\bm{C}=\bm{I}\big)\big]\big)$ & $\bm{Q}_{{\rm opt},\bm{X}_1}=\bm{V}_{\bm{A}_1}
  \bm{U}_{\rm DFT}^{\rm H}$ \\ \hline
 \textbf{Obj.\,4} & $f_{\text{A-Schur}}^{\text{Concave}}\big(\bm{d}\big[\bm{\Phi}_{\rm MSE}\big(\{\bm{F}_k\}_{k=1}^{K},
  \{\bm{Q}_{\bm{X}_k}\}_{k=1}^{K},\bm{C}=\bm{I}\big)\big]\big)$ & $\bm{Q}_{{\rm opt},\bm{X}_1}=\bm{V}_{\bm{A}_1}$ \\ \hline
 \textbf{Obj.\,5} & $f_{\text{M-Schur}}^{\text{Convex}}\big(\bm{d}\big[\bm{\Phi}_{\rm MSE}\big(\{\bm{F}_k\}_{k=1}^{K},
  \{\bm{Q}_{\bm{X}_k}\}_{k=1}^{K},\bm{C}\big)\big]\big)$ & $\bm{Q}_{{\rm opt},\bm{X}_1}=\bm{V}_{\bm{A}_1}
  \widetilde{\bm{U}}_{\rm GMD}^{\rm H}$ \\ \hline
 \textbf{Obj.\,6} & $f_{\text{M-Schur}}^{\text{Concave}}\big(\bm{d}\big[\bm{\Phi}_{\rm MSE}\big(\{\bm{F}_k\}_{k=1}^{K},
  \{\bm{Q}_{\bm{X}_k}\}_{k=1}^{K},\bm{C}\big)\big]\big)$ & $\bm{Q}_{{\rm opt},\bm{X}_1}=\bm{V}_{\bm{A}_1}$ \\ \hline
\end{tabular}
}
\end{center}
\label{Table_Objective_AF}
\vspace*{-4mm}
\end{table*}


 The $K$-hop AF MIMO relaying network is illustrated in Fig.~\ref{Fig5}, where the
 source, denoted as node~0, communicates with the destination, represented by
 node~$K$, with the help of the $(K-1)$ relays, which are nodes 1 to $(K-1)$. Denote
 the signal sent by the source as $\bm{x}_0$, which has the
 covariance matrix of $\sigma_{\bm{x}_0}^2\bm{I}$. Then the signal model in the
 $k$th hop, for $1\le k\le K$, can be expressed as
\begin{align}\label{eq125}
 \bm{x}_k =& \bm{H}_k \bm{X}_k \bm{x}_{k-1} + \bm{n}_k ,
\end{align}
 where $\bm{x}_k$ is the signal received by node $k$, $\bm{H}_k$ is the channel
 matrix of the $k$th hop, and $\bm{n}_k$ is the additive noise of the corresponding
 link with the covariance matrix $\sigma_{\bm{n}_k}^2\bm{I}$, while $\bm{X}_k$ is the
 forwarding matrix of node $(k-1)$. Note that $\bm{S}_1$ is the source's transmit precoding
 matrix. When the channel estimation error is considered, based on a practical channel estimation scheme \cite{Ding09} the CSI of the $k$th hop is
 expressed as
\begin{align}\label{eq126}
 \bm{H}_k =& \widehat{\bm{H}}_k + \bm{H}_{{\rm W},k} \bm{\Psi}_k^{\frac{1}{2}} ,
\end{align}
 where $\widehat{\bm{H}}_k$ and $\bm{H}_{{\rm W},k}\bm{\Psi}_k^{\frac{1}{2}}$ are the
 estimated CSI and the channel estimation error of the $k$th hop, respectively.
 Furthermore, $\bm{\Psi}_k$ is the covariance matrix of the channel estimate, and the
 elements of $\bm{H}_{{\rm W},k}$ follow the independent and identical complex Gaussian
 distribution $\mathcal{CN}(0,1)$. For notational convenience, let us define the new variables
$\bm{F}_1\bm{Q}_{\bm{X}_1}=\bm{X}_1$, with the associated unitary matrix $\bm{Q}_{\bm{X}_1}$, and
 $\bm{F}_k\bm{Q}_k$ for $2\le k\le K$ as
\begin{align}\label{F_definition} 
 \bm{F}_k =& \bm{X}_k \bm{K}_{{\rm{n}}_{k-1}}^{\frac{1}{2}} \bm{M}_{k-1} \bm{Q}_{\bm{X}_k}^{\rm H} ,
\end{align}
 where $\bm{Q}_k$ is the associated unitary matrix,
\begin{align}
 \bm{M}_{k}\! =& \Big(\! \bm{K}_{{\rm{n}}_{k}}^{-\frac{1}{2}}\widehat{\bm{H}}_{k}\bm{F}_{k}
  \bm{F}_{k}^{\rm H}\widehat{\bm{H}}_{k}^{\rm H}\bm{K}_{{\rm{n}}_{k}}^{-\frac{1}{2}}
  \! +\! \bm{I}\Big)^{\frac{1}{2}} \! ,\! \label{eq128} \\
  \bm{K}_{{\rm{n}}_{k}}\! =& \Big(\sigma_{\rm{n}_{k}}^2 + {\rm{Tr}}\big(\bm{F}_{k}
  \bm{F}_{k}^{\rm H}\bm{\Psi}_{k}\big) \Big) \bm{I} , \label{eq129}
\end{align}
 and clearly $\bm{K}_{\rm{n}_0}^{\frac{1}{2}}\bm{M}_0=\sigma_{\bm{x}_0}\bm{I}$. Based on these
 definitions, as proved in Appendix \ref{Appendix_MSE_Matrix}} the MSE matrix of the data detection at the destination is expressed as,
 \cite{XingTSP201502,XingTSP201601}
 \begin{align}\label{MSE_Matrix_M} 
 & \bm{\Phi}_{\rm MSE}\big(\{\bm{F}_k\}_{k=1}^{K},\{\bm{Q}_{\bm{X}_k}\}_{k=1}^{K},\bm{C}\big) \nonumber \\
 & \hspace*{5mm} = \sigma_{\bm{x}_0}^2 \bm{C} \bm{C}^{\rm H} - \sigma_{\bm{x}_0}^2 \bm{C}
  \left( \prod_{k=1}^{K} \bm{M}_k^{-\frac{1}{2}} \bm{K}_{{\rm{n}}_k}^{-\frac{1}{2}} \widehat{\bm{H}}_k
  \bm{F}_k \bm{Q}_{\bm{X}_k} \right)^{\rm H} \nonumber \\
 & \ \ \ \ \ \ \ \ \ \ \ \ \  \times\left( \prod_{k=1}^{K} \bm{M}_k^{-\frac{1}{2}} \bm{K}_{\rm{n}_k}^{-\frac{1}{2}}
  \widehat{\bm{H}}_k \bm{F}_k \bm{Q}_{\bm{X}_k} \right) \bm{C}^{\rm H} .
\end{align} Based on the MSE matrix given in
 (\ref{MSE_Matrix_M}), both the linear and nonlinear transceiver optimization problems
 \cite{XingTSP201502,XingTSP201601} can be unified into the general optimization
 problem \textbf{Opt.\,4.1} given in (\ref{Optimziation_Original}).  Various
 objective functions typically adopted for \textbf{Opt.\,4.1} are listed in
 Table~\ref{Table_Objective_AF}.   For linear transceiver optimization, to realize different
 levels of fairness between different transmitted data streams, a general objective
 function can be formulated as an additively Schur-convex function \cite{Majorization,XingTSP201601} or additively
 Schur-concave function \cite{Majorization,XingTSP201601} of the diagonal elements of the MSE matrix, which are given
 by \textbf{Obj.\,3} and \textbf{Obj.\,4} \cite{Majorization,XingTSP201601}, respectively.
 The additively
 Schur-convex function $f_{\text{A-Schur}}^{\rm convex}(\cdot )$ and the additively
 Schur-concave function $f_{\text{A-Schur}}^{\rm concave}(\cdot )$ represent different
 levels of fairness among the diagonal elements of the data MSE matrix. When nonlinear transceivers are adopted for improving the BER performance at the cost
 of increased complexity, e.g., THP or DFE, the objective functions of the transceiver
 optimization can be formulated as a multiplicative Schur-convex function or a
 multiplicative Schur-concave function of the vector consisting of the squared
 diagonal elements of the Cholesky-decomposition triangular matrix of the MSE matrix,
 that is, \textbf{Obj.\,5} and \textbf{Obj.\,6} \cite{Majorization,XingTSP201601}, respectively,  where $\bm{L}$ is
 a lower triangular matrix. The multiplicatively Schur-convex function
 $f_{\text{M-Schur}}^{\rm convex}(\cdot )$ and the multiplicatively Schur-concave function
 $f_{\text{M-Schur}}^{\rm concave}(\cdot )$ reflect the different levels of fairness
 among the different data streams, i.e., different tradeoffs among the performance of
 different data steams \cite{XingTSP201501}. The detailed definitions of $f_{\text{A-Schur}}^{\rm convex}(\cdot )$, $f_{\text{A-Schur}}^{\rm concave}(\cdot )$, $f_{\text{M-Schur}}^{\rm convex}(\cdot )$ and  $f_{\text{M-Schur}}^{\rm concave}(\cdot )$ are given in Appendix~\ref{Appendix_Majorization Theory}. This appendix makes our work self-contained.

The constraints $\psi_{k,i}(\bm{F}_k)\le 0$ are right
 unitarily invariant, and the power constraint model of \textbf{Opt.\,4.1} is more general
 than the power constraint models considered in \cite{XingTSP201501,XingTSP201502,XingTSP201601}.

 For linear transceivers with the objective functions \textbf{Objs.\,1-4} in
 Table~\ref{Table_Objective_AF}, $\bm{C}=\bm{I}$ is an identity matrix, while for nonlinear
 transceiver optimization with the objective functions \textbf{Obj.\,5} and \textbf{Obj.\,6} in
 Table~\ref{Table_Objective_AF} , $\bm{C}$ is a lower triangular matrix. Specifically, we assume
 that the size of $\bm{C}$ is $N\times N$. Then, for nonlinear transceivers, the
 optimal $\bm{C}$ satisfies \cite{XingTSP201601} \setcounter{equation}{75}
\begin{align}\label{eq132}
 \bm{C}_{\rm opt} =& {\rm diag} \big\{ \{ [\bm{L}]_{i,i} \}_{i=1}^N \big\} \bm{L}^{-1} ,
\end{align}
 where $\bm{L}$ is the triangular matrix of the Cholesky decomposition of the
 following matrix \cite{XingTSP201601}
\begin{align}\label{MSE_Matrix_M_A} 
 \bm{L} \bm{L}^{\rm H} =& \widetilde{\bm{\Phi}}_{\rm MSE}\big(\{\bm{F}_k\}_{k=1}^{K},
  \{\bm{Q}_{\bm{X}_k}\}_{k=1}^{K}\big) \nonumber \\
 =& \sigma_{\bm{x}_0}^2 \bm{I} - \sigma_{\bm{x}_0}^2
  \left( \prod_{k=1}^{K} \bm{M}_k^{-\frac{1}{2}} \bm{K}_{{\rm{n}}_k}^{-\frac{1}{2}} \widehat{\bm{H}}_k
  \bm{F}_k \bm{Q}_{\bm{X}_k} \right)^{\rm H} \nonumber \\
 & \hspace*{10mm} \times \left( \prod_{k=1}^{K} \bm{M}_k^{-\frac{1}{2}} \bm{K}_{{\rm{n}}_k}^{-\frac{1}{2}}
  \widehat{\bm{H}}_k \bm{F}_k \bm{Q}_{\bm{X}_k} \right).
\end{align}


The optimal unitary matrices $\bm{Q}_{\bm{X}_k}$ can be derived based on majorization theory.
Specifically, the optimal $\bm{Q}_k$ for $k > 1$ are derived as
 \cite{XingTSP201502,XingTSP201601,XingJSAC2012}
\begin{align}\label{eq134}
 \bm{Q}_{{\rm opt},\bm{X}_k} =& \bm{V}_{\bm{A}_k}\bm{U}_{\bm{A}_{k-1}}^{\rm H} ,
\end{align}
 where the unitary matrices $\bm{V}_{\bm{A}_k}$ and $\bm{U}_{\bm{A}_k}$ are defined by
 the following SVDs
\begin{align}\label{eq135}
 \bm{M}_k^{-\frac{1}{2}}\bm{K}_{{\rm{n}}_k}^{-\frac{1}{2}}\widehat{\bm{H}}_k\bm{F}_k =&
  \bm{U}_{\bm{A}_k}\bm{\Lambda}_{\bm{A}_k}\bm{V}_{\bm{A}_k}^{\rm H} \text{ with }
  \bm{\Lambda}_{\bm{A}_k} \searrow .
\end{align}
 The optimal $\bm{Q}_{\bm{X}_1}$ is determined by the specific objective function, and various
 $\bm{Q}_{{\rm opt},\bm{X}_1}$ associated with different objective functions are also
 summarized in Table~\ref{Table_Objective_AF}. Here, the unitary matrix
 $\bm{U}_{\rm{Arb}}$ denotes an arbitrary matrix having the appropriate dimension. The
 unitary matrix $\bm{U}_{\bm{W}}$ is the unitary matrix defined by the following EVD
\begin{align}\label{eq136}
 \bm{W} =& \bm{U}_{\bm{W}}\bm{\Lambda}_{\bm{W}}\bm{U}_{\bm{W}}^{\rm H} \text{ with }
  \bm{\Lambda}_{\bm{W}} \searrow .
\end{align}
 The unitary matrix $\bm{U}_{\rm DFT}$ is a DFT matrix \cite{XingTVT2016,Horn1990}. Finally, the unitary matrix
 $\widetilde{\bm{U}}_{\rm GMD}$ ensures that the triangular matrix of the Cholesky
 decomposition of $\widetilde{\bm{\Phi}}_{\rm MSE}\big(\{\bm{F}_k\}_{k=1}^{K},
 \{\bm{Q}_{\bm{X}_k}\}_{k=1}^{K}\big)$ has the same diagonal elements \cite{XingTSP201601}.

 Given the optimal $\bm{Q}_{{\rm opt},\bm{X}_k}$ and $\bm{C}_{\rm opt}$, the objective function
 of \textbf{Opt.\,4.1} can be rewritten as \cite{XingTSP201502}
 \begin{align}\label{eq137}
 & f\left( \bm{\Phi}_{\rm MSE}\big( \{\bm{F}_k\}_{k=1}^{K},\{\bm{Q}_{{\rm opt},\bm{X}_k}\}_{k=1}^{K},
  \bm{C}_{\rm opt}\big) \right) \nonumber \\
  = & \widetilde{f}\left( \left\{ \prod_{k=1}^{K} \frac{ \lambda_i(\bm{F}_k^{\rm H}
  \widehat{\bm{H}}_k^{\rm H} \bm{K}_{{\rm{n}}_k}^{-1} \widehat{\bm{H}}_k \bm{F}_k) }
  {1 + \lambda_i(\bm{F}^{\rm H}_k\widehat{\bm{H}}_k^{\rm H}\bm{K}_{{\rm{n}}_k}^{-1}
  \widehat{\bm{H}}_k\bm{F}_k)}\right\}_{i=1}^N \right) \nonumber \\
 \triangleq & f_{\text{Eigen}} \left(\left\{ \bm{\lambda}\big(\bm{F}_k^{\rm H}
  \widehat{\bm{H}}_k^{\rm H}\bm{K}_{{\rm{n}}_k}^{-1}\widehat{\bm{H}}_k\bm{F}_k\big)\right\}_{k=1}^K\right).
\end{align} In (\ref{eq137}) $f_{\text{Eigen}}(\cdot )$ is a monotonically decreasing function with respect
 to the eigenvalue vector $\bm{\lambda}\big(\bm{F}_k^{\rm H}\widehat{\bm{H}}_k^{\rm H}
 \bm{K}_{{\rm{n}}_k}^{-1}\widehat{\bm{H}}_k\bm{F}_k\big)$. The specific formula of $f_{\text{Eigen}}(\cdot )$ is determined by the specific performance metrics. For example, for sum MSE minimization $f_{\text{Eigen}}(\cdot )$ equals
 \begin{align}
 &f_{\text{Eigen}} \left(\left\{ \bm{\lambda}\big(\bm{F}_k^{\rm H}
  \widehat{\bm{H}}_k^{\rm H}\bm{K}_{{\rm{n}}_k}^{-1}\widehat{\bm{H}}_k\bm{F}_k\big)\right\}_{k=1}^K\right) \nonumber \\
  =&\sum_{i=1}^Ix_0^2\left(1- \prod_{k=1}^{K} \frac{ \lambda_i(\bm{F}_k^{\rm H}
  \widehat{\bm{H}}_k^{\rm H} \bm{K}_{\rm{n}_k}^{-1} \widehat{\bm{H}}_k \bm{F}_k) }
  {1 + \lambda_i(\bm{F}^{\rm H}_k\widehat{\bm{H}}_k^{\rm H}\bm{K}_{{\rm{n}}_k}^{-1}
  \widehat{\bm{H}}_k\bm{F}_k)}\right).
\end{align}In addition, for sum rate maximization $f_{\text{Eigen}}(\cdot )$ equals
\begin{align}
 &f_{\text{Eigen}} \left(\left\{ \bm{\lambda}\big(\bm{F}_k^{\rm H}
  \widehat{\bm{H}}_k^{\rm H}\bm{K}_{{\rm{n}}_k}^{-1}\widehat{\bm{H}}_k\bm{F}_k\big)\right\}_{k=1}^K\right) \nonumber \\
  =&\sum_{i=1}^I{\rm{log}}\left(1- \prod_{k=1}^{K} \frac{ \lambda_i(\bm{F}_k^{\rm H}
  \widehat{\bm{H}}_k^{\rm H} \bm{K}_{\rm{n}_k}^{-1} \widehat{\bm{H}}_k \bm{F}_k) }
  {1 + \lambda_i(\bm{F}^{\rm H}_k\widehat{\bm{H}}_k^{\rm H}\bm{K}_{{\rm{n}}_k}^{-1}
  \widehat{\bm{H}}_k\bm{F}_k)}\right).
\end{align}Hence, given $\bm{Q}_{{\rm opt},\bm{X}_k}$
 and $\bm{C}_{\rm opt}$, \textbf{Opt.\,4.1} is transferred into
\begin{align}\label{optimization_beginning} 
\hspace*{-2mm}\begin{array}{lcl}
 \textbf{Opt.\,4.2:}\!\! &\!\! \min\limits_{\{\bm{F}_k\}_{k=1}^{K}}\!\! &\!\! f_{\text{Eigen}}\left(
  \left\{ \bm{\lambda}\big(\bm{F}_k^{\rm H} \widehat{\bm{H}}_k^{\rm H}\bm{K}_{\rm{n}_k}^{-1}
  \widehat{\bm{H}}_k\bm{F}_k\big)\right\}_{k=1}^K\right) \! , \\
 \!\! &\!\! {\rm{s.t.}}\!\! &\!\! \bm{K}_{\rm{n}_k} = \Big(\sigma_{\rm{n}_k}^2 +
  {\rm{Tr}}\big(\bm{F}_k\bm{F}_k^{\rm H}\bm{\Psi}_k\big)\Big)\bm{I} , \\
 \!\! &\!\! \!\! &\!\! \psi_{k,i}(\bm{F}_k)\le 0 , 1\le i\le I_k, 1\le k\le K .
\end{array}
\end{align}

 Since the objective function of \textbf{Opt.\,4.2} is a monotonically decreasing
 function of $\bm{\lambda}\big(\bm{F}_k^{\rm H} \widehat{\bm{H}}_k^{\rm H}
 \bm{K}_{{\rm{n}}_k}^{-1}\widehat{\bm{H}}_k\bm{F}_k\big)$, it can be decoupled into the
 following sub-problems: for $1\le k\le K$,
\begin{align}\label{eq139}
\hspace*{-2mm}\begin{array}{lcl}
 \textbf{Opt.\,4.3:} & \min\limits_{\bm{F}_k} & \bm{\lambda}\big(\bm{F}_k^{\rm H} \widehat{\bm{H}}_k^{\rm H}
  \bm{K}_{\rm{n}_k}^{-1}\widehat{\bm{H}}_k\bm{F}_k\big) , \\
 & {\rm{s.t.}} & \bm{K}_{\rm{n}_k} = \Big(\sigma_{\rm{n}_k}^2\! +\!
  {\rm{Tr}}\big(\bm{F}_k\bm{F}_k^{\rm H}\bm{\Psi}_k\big)\Big) \bm{I} , \\
 & & \psi_{k,i}(\bm{F}_k)\le 0 , 1\le i\le I_k .
\end{array}
\end{align}
 Clearly, \textbf{Opt.\,4.3} is equivalent to the following matrix-monotonic
 optimization problem
\begin{align}\label{eq140}
\hspace*{-2mm}\begin{array}{lcl}
 \textbf{Opt.\,4.4:} & \min\limits_{\bm{F}_k} & \bm{F}_k^{\rm H} \widehat{\bm{H}}_k^{\rm H}
  \bm{K}_{\rm{n}_k}^{-1}\widehat{\bm{H}}_k\bm{F}_k , \\
 & {\rm{s.t.}} & \bm{K}_{\rm{n}_k} = \Big(\sigma_{\rm{n}_k}^2\! +\!
  {\rm{Tr}}\big(\bm{F}_k\bm{F}_k^{\rm H}\bm{\Psi}_k\big)\Big) \bm{I} , \\
 & & \psi_{k,i}(\bm{F}_k)\le 0 , 1\le i\le I_k .
\end{array}
\end{align}
 In this application, by exploiting its cascade structure, we are able to transfer
 the associated multiple-matrix-variate matrix-monotonic optimization problem into
  several single-matrix-variate matrix-monotonic optimization problems. Based
 on the fundamental results of the previous sections, we readily have the following
 results.

{\emph{1)}~{\rm\bf Shaping Constraint}}:  We have $I_k=1$ and
\begin{align}
 \psi_{k,1}(\bm{F}_k) =& \bm{F}_k\bm{F}_k^{\rm H} - \bm{R}_{{\rm s}_k} .
\end{align} As proved in Part I, based on Lemma~\ref{L1} in Section~\ref{S2}, it is concluded that when
 the rank of $\bm{R}_{{\rm s}_k}$ is not higher than the number of columns and the
 number of rows in $\bm{F}_k$, a suboptimal solution $\bm{F}_{{\rm opt},k}$ that maximizes a lower bound of the objective of  \textbf{Opt.\,4.4} is a
 square root of $\bm{R}_{{\rm s}_k}$. When $\bm{\Psi}_k=\bm{0}$ the lower bound is tight and then the suboptimal solution will be the Pareto optimal solution of \textbf{Opt.\,4.4}.

{\emph{2)}~{\rm\bf Joint Power Constraint}: We have
\begin{align}\label{eq119}
\begin{array}{l}
 \psi_{k,1}\big({\bm{F}}_k\big) = {\rm Tr}\big({\bm{F}}_k{\bm{F}}_k^{\rm H}\big) - P_k , \\
 \psi_{k,2}\big({\bm{F}}_k\big) = {\bm{F}}_k{\bm{F}}_k^{\rm H} - \tau_k \bm{I} .
\end{array}
\end{align}As proved in Part I, based on Lemma~2 in in Section~\ref{S2}  for the general case $\bm{\Psi}_k \not\propto \bm{I}$, a suboptimal solution that maximizes a lower bound of the objective of \textbf{Opt.\,4.4}
 satisfies the following structure
\begin{align}\label{eq141}
& \hspace*{-2mm}\bm{F}_k\hspace*{-1mm}=\hspace*{-1mm}\frac{\sigma_{{\rm{n}}_k}\widetilde{\bm{\Psi}}_k^{-\frac{1}{2}}
{\bm{V}}_{\widetilde{\bm{H}}_k}\bm{\Lambda}_{\widetilde{\bm{F}}_k}{\bm{U}}_{{\rm{Arb}},k}^{\rm{H}}}
{\left(1\hspace*{-1mm}-\hspace*{-1mm}
{\rm{Tr}}\left(\widetilde{\bm{\Psi}}_k^{-\frac{1}{2}}
\bm{\Psi}\widetilde{\bm{\Psi}}_k^{-\frac{1}{2}}
{\bm{V}}_{\widetilde{\bm{H}}_k}
\bm{\Lambda}_{\widetilde{\bm{F}}_k}\bm{\Lambda}_{\widetilde{\bm{F}}_k}^{\rm{H}}
{\bm{V}}_{\widetilde{\bm{H}}_k}^{\rm{H}}\right)\right)^{\frac{1}{2}}},
\end{align} where $\widetilde{\bm{\Psi}}_k=\sigma_{{\rm{n}}_k}^2\bm{I}
 +P_k\bm{\Psi}_k$. It is worth noting that when $\bm{\Psi}_k=\bm{0}$ or $\bm{\Psi}_k\propto \bm{I}$, the corresponding lower bound is tight. In other words, in that case the suboptimal solution is exactly the Pareto optimal solution of  \textbf{Opt.\,4.4}.
The unitary matrix $\bm{V}_{\widetilde{\bm{H}}_k}$ is the right unitary matrix of the following SVD
\begin{align}\label{eq142}
 \widehat{\bm{H}}_k\big(\sigma_{{\rm{n}}_k}^2\bm{I}+P_k\bm{\Psi}_k\big)^{-\frac{1}{2}}=
   {\bm{U}}_{\widetilde{\bm{H}}_k}\bm{\Lambda}_{\widetilde{\bm{H}}_k}{\bm{V}}_{\widetilde{\bm{H}}_k}^H, \text{with} \ \bm{\Lambda}_{\widetilde{\bm{H}}_k} \searrow ,
\end{align}
 and every diagonal element of the rectangular diagonal matrix ${{\bm{\Lambda}}}_{\widetilde{\bm{F}}_k}$ in (\ref{eq141})
 is smaller than the following threshold
 \begin{align}
\sqrt{\tau_k\big(\sigma_{{\rm{n}}_k}^2+P_k\lambda_{\min}(\bm{\Psi}_k)\big)/\big(\sigma_{{\rm{n}}_k}^2+P_k\lambda_{\max}(\bm{\Psi}_k)\big)}.
\end{align} The diagonal matrix $\bm{\Lambda}_{\widetilde{\bm{F}}_k}$ can be efficiently solved using a variant water-filling algorithm \cite{General_Waterfilling,Palomar_Waterfilling}.

{\emph{3)}~{\rm\bf Multiple weighted power constraints}: We have
\begin{align}
 \psi_{k,i}\big({\bm{F}}_k\big) =& {\rm Tr}\big(\bm{\Omega}_{k,i}{\bm{F}}_k
  {\bm{F}}_k^{\rm H}\big) - P_{k,i} .
\end{align}
 As proved in Part I, based on Lemma~\ref{L3} in Section~\ref{S2}, we conclude that the Pareto
 optimal solutions $\bm{F}_{{\rm opt},k}$ satisfy the following structure
\begin{align}\label{eq143}
 \bm{F}_{{\rm opt},k}\! =& \frac{\sigma_{{\rm{n}}_k} \widetilde{\bm{\Omega}}_k^{-\frac{1}{2}}
  \bm{V}_{\bm{\mathcal{H}}_k} \bm{\Lambda}_{\widetilde{\bm{F}}_k} \bm{U}_{\text{Arb},k}^{\rm H} }
  {\Big(\! 1\! -\! {\rm{Tr}}\big(\widetilde{\bm{\Omega}}_k^{-\frac{1}{2}} \bm{\Psi}_k
  \widetilde{\bm{\Omega}}_k^{-\frac{1}{2}} \bm{V}_{\bm{\mathcal{H}}_k}
  \bm{\Lambda}_{\widetilde{\bm{F}}_k}\bm{\Lambda}_{\widetilde{\bm{F}}_k}^{\rm H}
  \bm{V}_{\bm{\mathcal{H}}_k}^{\rm H} \big)\! \Big)^{\frac{1}{2}} }, \!
\end{align}
 where the unitary matrix $\bm{V}_{\bm{\mathcal{H}}_k}$ is defined by the SVD
\begin{align}\label{eq144}
 \widehat{\bm{H}}_k\widetilde{\bm{\Omega}}_k^{-\frac{1}{2}} =& \bm{U}_{\bm{\mathcal{H}}_k}
  \bm{\Lambda}_{\bm{\mathcal{H}}_k}\bm{V}_{\bm{\mathcal{H}}_k}^{\rm H} \text{ with }
  \bm{\Lambda}_{\bm{\mathcal{H}}_k} \searrow ,
\end{align}
 and the matrix $\widetilde{\bm{\Omega}}_k$ is defined by
\begin{align}
 \widetilde{\bm{\Omega}}_k =& \sigma_{\rm{n}_k}^2 \sum\nolimits_{i=1}^{I_k} \alpha_{k,i}
  \big( \bm{\Omega}_{k,i} + P_{k,i} \bm{\Psi}_k \big).
\end{align} The diagonal matrix $\bm{\Lambda}_{\widetilde{\bm{F}}_k}$ can be efficiently solved using water-filling algorithms \cite{Palomar_Waterfilling,General_Waterfilling}.

\section{Discussions}

In this paper, we have investigated three representative examples for the proposed framework of multi-variable matrix-monotonic optimization. Based on the proposed matrix-monotonic framework, the structure of the optimal solutions for the three largely different optimization problems can be derived in the same logic. The distinct difference between our work and existing work is that more general power constraints have been taken in account. Taking more general power constraints into account is definitely not trivial extensions. From physical meaning perspective, the considered optimization under more general power constraints includes more MIMO transceiver optimizations as its special cases. Moreover, from a mathematical viewpoint, the optimization with more general power constraints is more challenging. It is impossible to extend the existing results in the literature to the conclusions given in this paper via using simple substitutions. From convex optimization theory perspective, adding one more constraint may not change the convexity of the considered optimization problem. Specifically, adding one more linear matrix inequality on a SDP problem, the resulting problem is still a SDP problem. Adding a quadratical constraint on a QCQP problem, the resulting problem is still a QCQP. The story is totally different for the matrix-monotonic optimization framework as the matrix-monotonic optimization framework aims at deriving the structure of the optimal solutions. One more constraint will change the feasible region of matrix variate and significantly change the structure of the optimal solutions. The corresponding analytical derivations will change distinctly.

We also would like to point out that the matrix-monotonic optimization framework is applicable to more complicated communication systems. Recently, in \cite{XingTSP201901} based on the matrix-monotonic optimization framework, a general framework on hybrid transceiver optimizations under sum power constraint is proposed. Different from the fully digital MIMO systems, in a typical hybrid MIMO system, at the source or the destination the precoder or the receiver consists of two parts, i.e., analog part and digital part. For the analog part, only the phase of the signal at each antenna is adjustable. After that, in \cite{XingJSAC202001} based on the matrix-monotonic optimization framework, a framework on the transceiver optimizations for multi-hop AF hybrid MIMO relaying systems is further proposed. In multi-hop communications, the forwarding matrix at each relay consists of three parts, the left analog part, the inner digital part and the right analog part.

\section{Simulation Results and Discussions}\label{S9}

\subsection{Two-user MIMO Uplink}\label{S9.1}

 We first consider the MU-MIMO uplink,  where a pair of 4-antenna mobile users
 communicate with an 8-antenna BS. We define $\frac{P_k}{\sigma_n^2}$ as the SNR for the $k$th user, where $P_k$ is the sum transmit power of user
 $k$ and $\sigma_n^2$ is the noise power at each receive antenna of the BS. Without
 loss of generality, the same maximum transmit power is assumed for all the users, i.e.,
 $P_1=P_2$. Based on the Kronecker correlation model \cite{Ding09,Zhang2008,Jafar2005}, the spatial correlation matrix
 $\bm{R}_{\rm Rx}$ of the BS's receive antennas and the spatial correlation matrix
 $\bm{R}_{{\rm Tx},k}$ of the $k$th user's transmit antennas, where $k=1,2$, are
 specified respectively by $\big[\bm{R}_{\rm Rx}\big]_{i,j}=r_{\rm r}^{|i-j|}$ and
 $\big[\bm{R}_{{\rm Tx},k}\big]_{i,j}=r_{{\rm t},k}^{|i-j|}$. In the simulations, we
 further set $r_{{\rm t},1}=r_{{\rm t},2}=r_{{\rm t}}$. Three power constraints,
 namely, the shaping constraint, the joint power constraint and the per-antenna power
 constraints, are considered. For the shaping constraint, the widely used Kronecker
 correlation model of $\big[\bm{R}_{{\rm s}_k}\big]_{i,j}=0.6^{|i-j|}$ is employed \cite{XingTSP201502}. For the
 joint power constraint, the threshold is chosen as $\tau_k=1.4$. For the per-antenna
 power constraints, the power limits for the four antennas of each user are set to
 1.2, 1.2, 0.8 and 0.8, respectively.

\begin{figure}[!t]
\vspace*{-3mm}
\begin{center}
\includegraphics[width=0.48\textwidth]{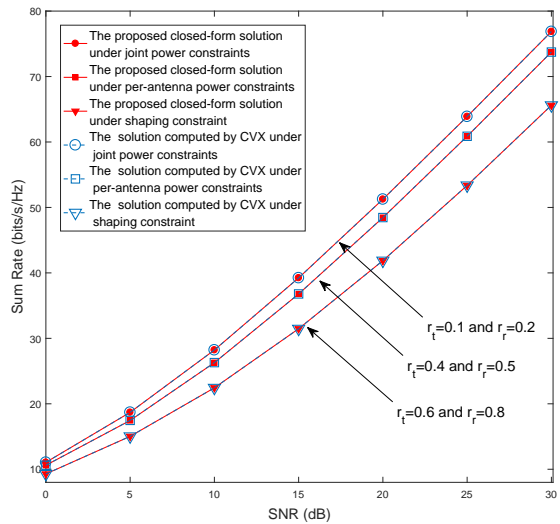}
\end{center}
\vspace*{-8mm}
\caption{Sum rate performance comparison between the proposed closed-form
 solutions and the solutions computed by the CVX tool for the two-user MIMO Uplink.}
\label{Fig6}
\vspace*{-4mm}
\end{figure}

 It is worth highlighting that the transceiver optimization under these three power
 constraints can be transferred into convex optimization problems, which can be solved
 numerically using the CVX tool \cite{tool}. This approach however suffers from high
 computational complexity, especially for high dimensional antenna arrays. By contrast,
 our approach presented in Section~\ref{S6} provides the optimal closed-form solutions
 for the same transceiver optimization design problems. Fig.~\ref{Fig6} compares the
 sum rate performance as the function of the SNR for the proposed closed-form solutions
 and for the numerical optimization solutions computed by the CVX tool. It can be seen that
 our closed-form solutions have an identical performance to the solutions computed by
 the CVX tool.

\subsection{Signal Compression for Distributed Sensor Networks}\label{S9.2}

In this subsection, we investigate the performance of the proposed
algorithm employed for signal compression in distributed sensor
networks. Specifically, the distributed sensor network considered
consists of $ K$ sensors and a data fusion center. Each sensor is
equipped with 4 antennas and the data fusion center is equipped with 8
antennas. The per-antenna power
constraints for the four antennas of each sensor are set to 1.2, 1.2,
0.8 and $ 0.8 $, respectively. For the signal correlations between
different sensors, the distance-dependent correlation matrix model
of~\cite{JFang2013} is adopted. Specifically, we have
$\bm{R}_{\bm{x}_{m,n}} = e^{-d_{m,n}}{\bm I}$ for the $m$th
sensor and the for the $n$th sensor, where $d_{m,n}$ is the
correlation between these two sensors. In our simulations, $d_{m,n}$
is distributed uniformly between 0 and 1. In order to quantify the
performance advantages attained, a benchmark algorithm based on CVX is
used in this subsection. The algorithm based on CVX aims for
minimizing the weighted sum MSE under per-antenna power constraints,
which is termed as the linear minimum mean square error (LMMSE)
algorithm. In the LMMSE algorithm, the signal compression matrices of
the different sensors and the combiner matrix at the data fusion
center are optimized iteratively. At each iteration, the optimization
problem considered is a standard QCQP problem, which can be readily
solved by CVX.  Observe in Fig.~\ref{Fig8} that the proposed algorithm
always outperforms the CVX-based benchmarker.

\begin{figure}[!t]
	\vspace*{-3mm}
	\begin{center}
		\includegraphics[width=0.48\textwidth]{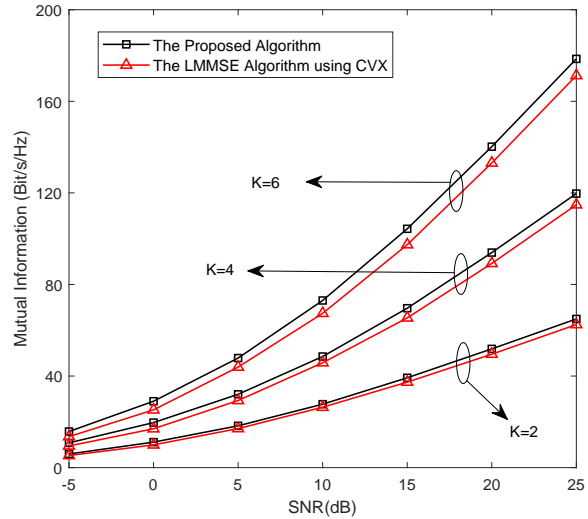}
	\end{center}
	\vspace*{-6mm}
	\caption{Mutual information performance comparisons between the proposed algorithm and the LMMSE algorithm based on CVX for distributed sensor networks with different numbers of sensors.}
	\label{Fig8}
	\vspace*{-4mm}
\end{figure}

\subsection{Dual-hop AF MIMO Relaying Network}\label{S9.2}

 A dual-hop AF MIMO relaying network is simulated, which consists of one source, one
 relay and one destination. All the nodes are equipped with 4 antennas. At the source
 and relay, per-antenna power constraints are imposed. Specifically, the power limits
 for the four antennas are set as 1, 1, 1 and 1, respectively. The SNR in each hop is
 defined as the ratio between the transmit power and the noise variance, i.e.,
 $\text{SNR}_k=\frac{P_k}{\sigma_{{\rm{n}}_k}^2}$. Without loss of generality, the SNRs in the
 both hops are assumed to be the same, namely, $\text{SNR}_1=\text{SNR}_2=\text{SNR}$.

 In contrast to the existing works \cite{XingTSP201502,XingTSP201601}, which consider
 the transceiver optimization unrealistically with the perfect CSI, in this paper, we
 focus on the robust transceiver optimization, which takes into account the channel
 estimation error. In the simulations, the estimated channel matrix is generated
 according to $\widehat{\bm{H}}_k=\widehat{\bm{H}}_{{\rm{W}},k}\bm{\Psi}_k^{\frac{1}{2}}$
 \cite{XingTSP201501}, where we have $\big[\bm{\Psi}_k\big]_{i,j}=0.6^{|i-j|}$. The elements
 of $\widehat{\bm{H}}_{{\rm{W}},k}$ are independently identically distributed Gaussian
 random variables. In order to ensure that $\mathbb{E}\big\{\big[\bm{H}\big]_{i,j}
 \big[\bm{H}\big]_{i,j}^*\big\}=1$, $\forall i,j$, we set $\mathbb{E}\big\{
 \big[\bm{H}_{{\rm W},k}\big]_{i,j}\big[\bm{H}_{{\rm W},k}\big]_{i,j}^*\big\}=
 \sigma_{e_k}^2$ and $\mathbb{E}\big\{\big[\widehat{\bm{H}}_{{\rm W},k}\big]_{i,j}
 \big[\widehat{\bm{H}}_{{\rm W},k}\big]_{i,j}^*\big\}=1-\sigma_{e_k}^2$. Without loss
 of generality, we assume $\sigma_{e_1}^2=\sigma_{e_2}^2=\sigma_e^2$. It
 can be seen from Fig.~\ref{Fig7} that our robust design achieves better sum rate performance than the
 non-robust design of \cite{XingTSP201601}. Furthermore, as expected, the performance
 gap between the robust and non-robust designs becomes larger as the channel
 estimation error increases.

\begin{figure}[!t]
\vspace*{-3mm}
\begin{center}
\includegraphics[width=0.48\textwidth]{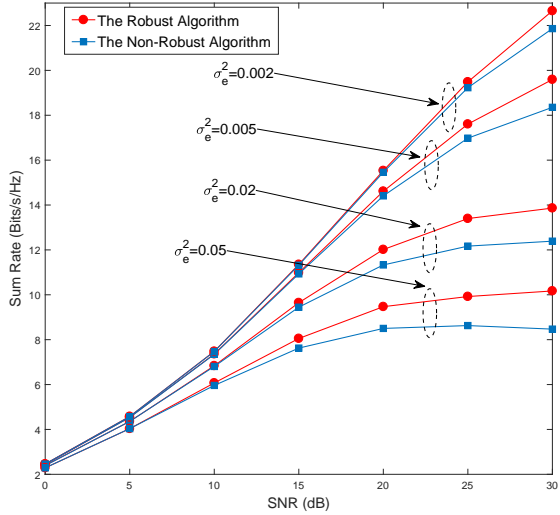}
\end{center}
\vspace*{-7mm}
\caption{Sum rate performance comparison between our proposed robust design and the
 non-robust design of \cite{XingTSP201601} for the dual-hop AF MIMO
 relaying network.}
\label{Fig7}
\vspace*{-4mm}
\end{figure}
\section{Conclusions}\label{S10}

In this paper, we investigated the application of the framework of matrix-monotonic optimization in the optimizations with multiple matrix-variates. It is shown that when several properties are satisfied, the framework of matrix-monotonic optimization still works, based on which the optimal structures of multiple matrix-variates can be derived. Then the multiple matrix-variable optimizations can be effectively solved in iterative manners. Three specific examples are also given in this paper to verify the validity of the proposed multi-variable matrix-monotonic optimization framework. Specifically, under various power constraints, i.e., sum power constraint, shaping constraints, joint power constraints and multiple weighted power constraints,
 the transceiver optimizations for uplink MIMO communications, the compression matrix optimizations for distributed sensor networks, and the robust transceiver optimizations for multi-hop AF MIMO relaying systems have been investigated. At the end of this paper, several numerical results demonstrated the accuracy and performance advantages of the proposed  multi-variable matrix-monotonic optimization framework.

\appendices

\section{Computation of $\bm{P}_k$ and $\bm{\Xi}_k$}
\label{Appendix_Sensor_Precoding}

Given the following block diagonal matrix
\begin{align}\label{Phi}
\bm{\Phi}={\rm diag}\Big\{ \big\{\bm{X}_k^{\rm H}\bm{H}_k^{\rm H}\bm{R}_{\rm{n}_k}^{-1}\bm{H}_k
  \bm{X}_k\big\}_{k=1}^K\Big\}
\end{align} the permutation matrix $\bm{P}_k$ aims at changing the orders of the $k$th element $\bm{X}_k^{\rm H}\bm{H}_k^{\rm H}\bm{R}_{\rm{n}_k}^{-1}\bm{H}_k
  \bm{X}_k$ and the first element $\bm{X}_1^{\rm H}\bm{H}_1^{\rm H}\bm{R}_{\rm{n}_1}^{-1}\bm{H}_1
  \bm{X}_1$ along the diagonal line. Before constructing $\bm{P}$, we first give an identity matrix $\bm{I}$ that has the same dimensions as $\bm{\Phi}$. Moreover,  $\bm{I}$ can be interpreted as a block diagonal matrix as
\begin{align}
{\bm{I}}={\rm{diag}}\Big\{ \big\{\bm{I}_k\big\}_{k=1}^K\Big\}
\end{align}where $\bm{I}_k$  is an identity matrix of the same dimensions as $\bm{X}_k^{\rm H}\bm{H}_k^{\rm H}\bm{R}_{\rm{n}_k}^{-1}\bm{H}_k
  \bm{X}_k$ for $1 \le k \le K$. Moreover, $\bm{I}$ is further divided into the following submatrices
\begin{align}
\label{Index_Equ}
{\bm{I}}={\rm{diag}}\Big\{ \big\{\bm{I}_k\big\}_{k=1}^K\Big\}=\left[ {\begin{array}{*{20}{c}}
{{\bm{\mathcal{I}}}_1}\\
{{\bm{\mathcal{I}}}_2}\\
{\vdots}\\
{{\bm{\mathcal{I}}}_K}
\end{array}} \right]
\end{align} where ${\bm{\mathcal{I}}}_k$ and ${\bm{I}}_k$ have the same row number for $1 \le k \le K$.
Based on the above definitions of ${\bm{\mathcal{I}}}_k$'s, we have
\begin{align}\label{property_1}
{\bm{\mathcal{I}}}_k\bm{\Phi}{\bm{\mathcal{I}}}_j^{\rm{H}}= {\bm{\mathcal{I}}}_k\bm{\Phi}{\bm{\mathcal{I}}}_j^{\rm{T}}={\bm{0}}, {\rm{for}}, k\not= j,
\end{align} and
\begin{align}\label{property_2}
{\bm{\mathcal{I}}}_k\bm{\Phi}{\bm{\mathcal{I}}}_k^{\rm{H}}={\bm{\mathcal{I}}}_k\bm{\Phi}
{\bm{\mathcal{I}}}_k^{\rm{T}}=\bm{X}_k^{\rm H}\bm{H}_k^{\rm H}\bm{R}_{\rm{n}_k}^{-1}\bm{H}_k
  \bm{X}_k.
\end{align} Therefore, based on (\ref{Index_Equ}) $\bm{P}_k$ is constructed by interchanging ${\bm{\mathcal{I}}}_1$ and ${\bm{\mathcal{I}}}_k$, i.e.,
\begin{align}
\label{P_Equ}
\bm{P}_k=\left[ {\begin{array}{*{20}{c}}
{{\bm{\mathcal{I}}}_k}\\
{{\bm{\mathcal{I}}}_2}\\
{\vdots}\\
{{\bm{\mathcal{I}}}_{k-1}}\\
{{\bm{\mathcal{I}}}_1}\\
{{\bm{\mathcal{I}}}_{k+1}}\\
{\vdots}\\
{{\bm{\mathcal{I}}}_K}
\end{array}} \right].
\end{align} It is obvious that $\bm{P}_k$ is a unitary matrix, i.e.,
\begin{align}
\bm{P}_k\bm{P}_k^{\rm{H}}=\bm{I} \ \text{and} \ \bm{P}_k^{\rm{H}}\bm{P}_k=\bm{I}.
\end{align}
Based on (\ref{P_Equ}) and together with (\ref{property_1}) and (\ref{property_2}), we have
\begin{align}
 & \bm{P}_k{\rm diag}\Big\{ \big\{\bm{X}_k^{\rm H}\bm{H}_k^{\rm H}\bm{R}_{{\rm{n}}_k}^{-1}\bm{H}_k
  \bm{X}_k\big\}_{k=1}^K\Big\} \bm{P}_k^{\rm H} \nonumber \\
 =&\left[ \begin{array}{cc}
  \bm{X}_k^{\rm H}\bm{H}_k^{\rm H}\bm{R}_{\rm{n}_k}^{-1}\bm{H}_k \bm{X}_k & \bm{0} \\
  \bm{0} & \bm{\Xi}_k \end{array} \right].
\end{align} where $\bm{\Xi}_k$ is the following block diagonal matrix
\begin{align}
\bm{\Xi}_k={\rm{diag}}\{\widetilde{\bm{\Phi}}_{2},\cdots,\widetilde{\bm{\Phi}}_{k-1},
\widetilde{\bm{\Phi}}_{1},
\widetilde{\bm{\Phi}}_{k+1},\cdots,\widetilde{\bm{\Phi}}_{K}\}
\end{align} with $\widetilde{\bm{\Phi}}_{j}=\bm{X}_j^{\rm H}\bm{H}_j^{\rm H}\bm{R}_{\rm{n}_j}^{-1}\bm{H}_j \bm{X}_j$.

\section{MSE Matrix for Multi-Hop Communications}
\label{Appendix_MSE_Matrix}

Based on the signal model given in (\ref{eq125}), at the destination the received signal $\bm{y}$ equals
\begin{align}\label{app_signal_model}
\bm{y}=\bm{x}_K =& \bm{H}_K \bm{X}_K \bm{x}_{K-1} + \bm{n}_K.
\end{align} After performing a linear equalizer $\bm{G}$, the signal estimation MSE matrix at the destination can be written in the following formula \cite{XingJSAC2012,XingTSP201601,XingTSP201502}
\begin{align}
\label{app_MSE_Matrix_0}
&\bm{\Phi}_{\rm MSE}\big(\bm{G},\{\bm{X}_k\}_{k=1}^{K},\bm{C}\big)\nonumber \\
&=\mathbb{E}\{({\bm{G}}{\bm{y}}-\bm{C}\bm{x}_0)
({\bm{G}}{\bm{y}}-\bm{C}\bm{x}_0)^{\rm{H}}\}
\end{align}where $\bm{C}=\bm{I}+\bm{B}$ and $\bm{B}$ is a strictly lower triangular matrix \cite{XingTSP201501}. For the linear transceivers, $\bm{B}$ is a constant matrix, i.e., $\bm{B}=\bm{0}$.
On the other hand for the nonlinear transceivers with THP or DFE, $\bm{B}$ corresponds to the feedback operations and should  be optimized as well \cite{XingJSAC2012,XingTSP201601,XingTSP201502}. Substituting (\ref{app_signal_model}) into $\bm{\Phi}_{\rm MSE}\big(\bm{G},\{\bm{X}_k\}_{k=1}^{K}\big)$ in (\ref{app_MSE_Matrix_0}), we have
\begin{align}\label{app_MSE_Matrix}
&\bm{\Phi}_{\rm MSE}\big(\bm{G},\{\bm{X}_k\}_{k=1}^{K},\bm{C}\big)\nonumber\\
=&\bm{G}\left(\bm{\widehat H}_K \bm{X}_K  \bm{R}_{\bm{x}_{K-1}}\bm{X}_K^{\rm{H}} \bm{\widehat H}_K^{\rm{H}}+{\rm{Tr}}(\bm{X}_K  \bm{R}_{\bm{x}_{K-1}}\bm{X}_K^{\rm{H}} {\boldsymbol{\Psi}}_K)\bm{I}\right) \bm{G}^{\rm{H}}\nonumber \\
&-\bm{G}\left(\prod_{k=1}^{K} \bm{\widehat H}_k \bm{X}_k\right) \bm{R}_{\bm{x}_{0}}\bm{C}^{\rm{H}}-\bm{C}\bm{R}_{\bm{x}_{0}} \left(\prod_{k=1}^{K} \bm{\widehat H}_k \bm{X}_k\right)^{\rm{H}} \bm{G}^{\rm{H}}\nonumber \\
&+\bm{G} \bm{R}_{\bm{n}_K} \bm{G}^{\rm{H}}+\bm{C}\bm{R}_{\bm{x}_{0}}\bm{C}^{\rm{H}},
\end{align}where $\bm{R}_{\bm{x}_k}=\mathbb{E}\{\bm{x}_k\bm{x}_k^{\rm{H}}\}$. The corresponding LMMSE equalizer $\bm{G}_{\rm{LMMSE}}$ equals
\begin{align}\label{G_LMMSE}
\bm{G}_{\rm{LMMSE}}
=  \ &\bm{C}\bm{R}_{\bm{x}_{0}}\left(\prod_{k=1}^{K} \bm{\widehat H}_k \bm{X}_k\right)^{\rm{H}}\nonumber \\
&\times\left(\bm{\widehat H}_K \bm{X}_K  \bm{R}_{\bm{x}_{K-1}}\bm{X}_K^{\rm{H}} \bm{\widehat H}_K^{\rm{H}}+\bm{K}_{{\rm{n}}_K}\right)^{-1}
\end{align}with
\begin{align}
\bm{K}_{{\rm{n}}_K}={\rm{Tr}}(\bm{X}_K  \bm{R}_{\bm{x}_{K-1}}\bm{X}_K^{\rm{H}} {\boldsymbol{\Psi}}_K)\bm{I}+ \bm{R}_{\bm{n}_K} .
\end{align} It is well-known that the LMMSE equalizer $\bm{G}_{\rm{LMMSE}}$ is the optimal $\bm{G}$ for  $\bm{\Phi}_{\rm MSE}\big(\bm{G},\{\bm{X}_k\}_{k=1}^{K}\big)$ as \cite{XingTSP201501}
\begin{align}
\bm{\Phi}_{\rm MSE}\big(\bm{G},\{\bm{X}_k\}_{k=1}^{K},\bm{C}\big) \succeq \bm{\Phi}_{\rm MSE}\big(\bm{G}_{\rm{LMMSE}},\{\bm{X}_k\}_{k=1}^{K},\bm{C}\big).
\end{align} Substituting $\bm{G}_{\rm{LMMSE}}$ into (\ref{app_MSE_Matrix}), we have
\begin{align}
&\bm{\Phi}_{\rm MSE}\big(\bm{G},\{\bm{X}_k\}_{k=1}^{K},\bm{C}\big)\nonumber\\
&=\bm{C}\bm{R}_{\bm{x}_{0}}\bm{C}^{\rm{H}}-\bm{C}\bm{R}_{\bm{x}_{0}}\left(\prod_{k=1}^{K} \bm{\widehat H}_k \bm{X}_k\right)^{\rm{H}}\nonumber \\
&  \ \ \ \ \ \ \ \ \ \ \ \times\left(\bm{\widehat H}_K \bm{X}_K  \bm{R}_{\bm{x}_{K-1}}\bm{X}_K^{\rm{H}} \bm{\widehat H}_K^{\rm{H}}+\bm{K}_{{\rm{n}}_K}\right)^{-1}\nonumber \\
& \ \ \ \ \ \ \ \ \ \ \  \ \ \ \ \ \ \ \ \ \ \  \ \ \ \ \ \ \    \times\left(\prod_{k=1}^{K} \bm{\widehat H}_k \bm{X}_k\right)\bm{R}_{\bm{x}_{0}}\bm{C}^{\rm{H}}.
\end{align} Therefore, based on the definition of $\bm{F}_{k}$ in (\ref{F_definition}) and the definition of $\bm{M}_k$ in (\ref{eq128})  we have
\begin{align}
&\bm{\Phi}_{\rm MSE}\big(\bm{G},\{\bm{X}_k\}_{k=1}^{K},\bm{C}\big)\nonumber\\
=\ & \bm{\Phi}_{\rm MSE}\big(\{\bm{F}_k\}_{k=1}^{K},\{\bm{Q}_{\bm{X}_k}\}_{k=1}^{K},\bm{C}\big) \nonumber \\
= \ & \sigma_{\bm{x}_0}^2 \bm{C} \bm{C}^{\rm H} - \sigma_{\bm{x}_0}^2 \bm{C}
  \left( \prod_{k=1}^{K} \bm{M}_k^{-\frac{1}{2}} \bm{K}_{{\rm{n}}_k}^{-\frac{1}{2}} \widehat{\bm{H}}_k
  \bm{F}_k \bm{Q}_{\bm{X}_k} \right)^{\rm H} \nonumber \\
 & \ \ \ \ \ \ \ \ \ \ \ \ \  \times\left( \prod_{k=1}^{K} \bm{M}_k^{-\frac{1}{2}} \bm{K}_{\rm{n}_k}^{-\frac{1}{2}}
  \widehat{\bm{H}}_k \bm{F}_k \bm{Q}_{\bm{X}_k} \right) \bm{C}^{\rm H} .
\end{align}

\section{Fundamental Definitions of Majorization Theory}
\label{Appendix_Majorization Theory}

A brief introduction of majorization theory is given in this appendix. Generally speaking, majorization theory is an important branch of matrix inequality theory \cite{Marshall79}. Majorization theory is a very useful mathematical tool to prove the inequalities for the diagonal elements of matrices, the eigenvalues of matrices and the singular values of matrices.  Majorization theory can reveal the relationships between diagonal elements and eigenvalues, based on which some extrema can be computed. Moreover, majorization theory can quantitatively analyze the relationships between the eigenvalues or singular values of matrix products and matrix additions and that of the involved individual matrices. Based on majorization theory, a rich body of useful matrix inequalities can be derived, based on which the extrema of the matrix variate functions can be derived.
The definitions of additively Schur-convex, additively Schur-concave, multiplicatively Schur-convex and multiplciatively Schur-concave functions are given in the following. Meanwhile, we would like to point out that Schur-convex function is a kind of increasing function and Schur-concave function is a kind of decreasing function \cite{XingTSP201502}. They actually have no relationship with the traditional convex or concave properties defined in the convex optimization theory \cite{Boyd04}.

\begin{definition} [\!\!\cite{Marshall79}]
For a $K\times 1$ vector $ {\bm x} \in \mathbb{R}^{K} $, the $ \ell $th largest element of $ {\bm x} $ is denoted as $ {x}_{[\ell]} $, i.e., $ {x}_{[1]} \ge {x}_{[2]} \ge \cdots \ge {x}_{[K]} $. Based on this definition, for two $K\times 1$ vectors $ {\bm x}, {\bm y} \in \mathbb{R}^{K} $, the statement that $ {\bm y} $ majorizes $ {\bm x} $ additively, denoted by $ {\bm x} \prec_{+} {\bm y} $, is defined as follows
\begin{align}
&\sum_{n = 1}^{m} {x}_{[n]} \le \sum_{n = 1}^{m} {y}_{[n]}, \; m = 1,\cdots, K\hspace*{-1mm}-\hspace*{-1mm}1, \text{and} \sum_{n = 1}^{K} {x}_{[n]} = \sum_{n = 1}^{K} {y}_{[n]}.
\end{align}\end{definition}

\begin{definition} [\!\!\cite{Marshall79}]
A real function $ f(\cdot) $ is additively Schur-convex when the following relationship holds
\begin{align}
 f( {\bm x} ) \le f( {\bm y} ) \ \text{when} \ {\bm x} \prec_{+} {\bm y}.
\end{align} A real function $ f(\cdot)$ is additively Schur-concave if and only if $ -f(\cdot) $ is additively Schur-convex.\end{definition}

\begin{definition} [\!\!\cite{XingTSP201502,Majorization}]
 Given $K\times 1$ vectors $ {\bm x}, {\bm y} \in \mathbb{R}^{K} $ with nonnegative elements, the statement that the vector ${\bm y} $ majorizes vector $ {\bm x} $ multiplicatively, denoted by $ {\bm x} \prec_{\times} {\bm y} $, is defined as follows
\begin{align}
\prod_{n = 1}^{m} {x}_{[n]} \le \prod_{n = 1}^{m} {y}_{[n]}, \; m = 1,\cdots, K\hspace*{-1mm}-\hspace*{-1mm}1, \text{and}
 \prod_{n = 1}^{K} {x}_{[n]} = \prod_{n = 1}^{K} {y}_{[n]}.
\end{align}\end{definition}

\begin{definition} [\!\!\cite{XingTSP201502,Majorization}]
A real function $ f(\cdot) $ is multiplicatively Schur-convex when the following relationship holds
\begin{align}
 f( {\bm x} ) \le f( {\bm y} ) \ \text{when} \  {\bm x} \prec_{\times} {\bm y }.
\end{align} A real function $ f(\cdot) $ is multiplicatively Schur-concave if and only if $ -f(\cdot) $ is multiplicatively Schur-convex.\end{definition}

Generally, it is not convenient to use these definitions to prove whether a function is Schur-convex or not. In the following, two criteria are given, based on which we can judge whether a function is additively Schur-convex or multiplicatively Schur-convex \cite{XingJSAC2012,XingTSP201501,XingTSP2013} .
For a given function $f(\cdot)$, according to the value order of the elements of $\bm{x}$ the considered function $f(\bm{x})$ is first reformulated as
\begin{align}
f(\bm{x})=\psi(x_{[1]},\cdots,x_{[k]},x_{[k+1]},\cdots).
\end{align} When $f(\bm{x})=\psi(x_{[1]},\cdots,x_{[k]}-e,x_{[k+1]}+e,\cdots)$ is a decreasing function with respect to $e$ for $e\ge0$ and $x_{[k]}-e\ge x_{[k+1]}+e$, $f(\cdot)$ is additively Schur-convex. On the other hand, when $f(\bm{x})=\psi(x_{[1]},\cdots,x_{[k]}/e,x_{[k+1]}e,\cdots)$ is a decreasing function with respect to $e$ for $e\ge1$ and $x_{[k]}/e\ge x_{[k+1]}e$, $f(\cdot)$ is multiplicatively Schur-convex.

\end{document}